\documentclass[prd,eqsecnum,twocolumn,amsfonts,amssymb]{revtex4}

\usepackage{graphicx}

\usepackage{bm}

\newcommand{\beq}{\begin{equation}}
\newcommand{\eeq}{\end{equation}}
\newcommand{\beqs}{\begin{eqnarray}}
\newcommand{\eeqs}{\end{eqnarray}}
\newcommand{\lsim}{\mathrel{\raisebox{-
.6ex}{$\stackrel{\textstyle<}{\sim}$}}}

\newcommand{\drawsquare}[2]{\hbox{%
\rule{#2pt}{#1pt}\hskip-#2pt
\rule{#1pt}{#2pt}\hskip-#1pt
\rule[#1pt]{#1pt}{#2pt}}\rule[#1pt]{#2pt}{#2pt}\hskip-#2pt
\rule{#2pt}{#1pt}}
\newcommand{\asym}{\raisebox{-3.5pt}{\drawsquare{6.5}{0.4}}\hskip-6.9pt%
        \raisebox{3pt}{\drawsquare{6.5}{0.4}}}

\begin{document}

\title{Generational Structure of Models with Dynamical Symmetry Breaking} 

\author{Thomas A. Ryttov}

\author{Robert Shrock}

\affiliation{
C. N. Yang Institute for Theoretical Physics \\
Stony Brook University \\
Stony Brook, NY 11794}

\begin{abstract}

In models with dynamical electroweak symmetry breaking, this breaking is
normally communicated to quarks and leptons by a set of vector bosons with
masses generated via sequential breaking of a larger gauge symmetry. In
reasonably ultraviolet-complete theories of this type, the number of stages of
breaking of the larger gauge symmetry is usually equal to the observed number
of quark and lepton generations, $N_{gen.}=3$. Here we investigate the general
question of how the construction and properties of these models depend on
$N_{gen.}$, regarded as a variable. We build and analyze models with
illustrative values of $N_{gen.}$ different from 3 (namely, $N_{gen.}=1,2,4$)
that exhibit the necessary sequential symmetry breaking down to a strongly
coupled sector that dynamically breaks electroweak symmetry.  Our results for
variable $N_{gen.}$ show that one can robustly obtain, for this latter sector,
a theory with a gauge coupling that is large but slowly running, controlled by
an approximate infrared fixed point of the renormalization group.  Owing to
this, we find that for all of the values of $N_{gen.}$ considered,
standard-model fermions of the highest generation have masses that can be
comparable to the electroweak-symmetry breaking scale. We also study the
interplay of multiple strongly coupled gauge symmetries in these models.

\end{abstract}


\maketitle

\section{Introduction} 
\label{intro}

Electroweak symmetry breaking (EWSB) plays a crucial role for observed particle
interactions, but its origin remains an outstanding mystery. There are actually
several aspects to this physics; in addition to a mechanism to explain how the
$W$ and $Z$ bosons acquire masses, there is also the necessity to explain how
the quarks and charged leptons gain masses, why they come in three generations,
and why their masses exhibit the generational hierarchy that they do.
Explaining the very small neutrino masses is yet another challenge.  In an
appealing class of theories with dynamical electroweak symmetry breaking, this
breaking is produced by means of an asymptotically free, vectorial, gauge
interaction based on an exact gauge symmetry, commonly called technicolor (TC),
that becomes strongly coupled on the TeV scale, causing the formation of
bilinear technifermion condensates \cite{tc}.  In these theories, the EWSB is
communicated to the standard-model (SM) fermions, which are technisinglets, via
exchanges of massive gauge bosons associated with a higher symmetry, extended
technicolor (ETC) \cite{etc}.  Some early related works on gauge symmetry
breaking include \cite{early}; some recent reviews of TC/ETC theories include
\cite{nag06}-\cite{sanrev}.

An important aspect of such theories is the pattern of sequential extended
technicolor symmetry breaking to the residual technicolor symmetry, since this
determines the hierarchical generational mass spectrum of the standard-model
fermions. Early works tended to model ETC effects via (non-renormalizable)
four-fermion operators connecting SM fermions and technifermions, with some
assumed values for their coefficients.  More complete studies took on the task
of deriving these four-fermion operators by analyses of renormalizable,
reasonably ultraviolet-complete, ETC theories.  In particular, detailed studies
were carried out for reasonably ultraviolet-complete ETC models containing an
SU(2)$_{TC}$ technicolor gauge group, with ETC symmetry-breaking patterns
giving rise to the observed three generations \cite{ngen3} of standard-model
fermions \cite{at94} and to acceptably light neutrinos \cite{nt}-\cite{kt},
\cite{nag06}. As is evident from these models, there is a tight connection
between the number of standard-model fermion generations and the
sequential breaking of the ETC symmetry down to the technicolor subgroup.

The studies of reasonably ultraviolet-complete TC/ETC models naturally lead one
to investigate a more general and abstract topic, namely the connection between
the properties of the enveloping ETC theory and the number of standard-model
fermion generations, $N_{gen.}$, when $N_{gen.}$ is taken as a variable rather
than being fixed at its inferred physical value of 3 \cite{ngen3}. We address
this question here, considering the hypothetical values $N_{gen.}=1, \ 2$, and
4.  The purpose of our analysis is not to try to produce a quasi-realistic ETC
model, but instead to investigate how the value of $N_{gen.}$ influences the
construction and properties of the model. There are several interesting
questions that one can investigate in this context. In several previous
detailed studies of reasonably ultraviolet-complete quasi-realistic TC/ETC
models \cite{at94}, \cite{nt}-\cite{kt}, one relied upon an auxiliary strongly
coupled gauge symmetry, called hypercolor (HC), to produce the requisite
sequential ETC symmetry breaking. A natural question to ask is whether, for
values of $N_{gen.}$ different from 3, in particular, for the apparently
simpler cases $N_{gen.}=1$ or $N_{gen.}=2$, one might be able to construct a
TC/ETC theory in which all of the ETC symmetry breaking could be accomplished
by strong self-breaking of the ETC gauge symmetry, without the aid of this
auxiliary strongly coupled HC gauge interaction.  A second and related topic
for investigation is the evolution of the ETC and HC interactions from high
energy scales down to lower ones, through the various sequential breakings of
the high-scale ETC symmetry.  This entails an examination of the effective
field theories that are operative at the different scales, including their
content of dynamical fermions and their plausible channels of condensation.
One aspect of this is to check that the relevant strongly coupled gauge
interactions do, indeed, plausibly produce the requisite bilinear fermion
condensates for ETC symmetry breaking and, in the case of the residual TC
theory, the technifermion condensates that cause dynamical electroweak symmetry
breaking, rather than evolving into the infrared in a chirally symmetric
manner, with unwanted non-Abelian Coulombic behavior. Yet another interesting
question concerns the effect of changing the value of $N_{gen.}$ on the
properties of the residual technicolor theory, and whether this TC theory can
easily have a large but slowly running gauge coupling associated with an
approximate infrared fixed point. A profound mystery pertaining to the observed
quarks and leptons is why all of them except for one -- the top quark -- have
masses that are considerably smaller than the electroweak symmetry-breaking
scale.  Thus, a final question concerns what generic predictions these models
with other values of $N_{gen.}$ make about the standard-model quarks and
leptons.  Our illustrative models with variable $N_{gen.} \ne 3$ provide a
useful theoretical laboratory in which to investigate these questions.

Although our main focus here is on the general field-theoretic question of the
role of $N_{gen.}$ in TC/ETC model-building, we note that there is currently
continuing interest in the possibility that there really are four (or more)
generations of SM fermions.  Reviews include Refs. \cite{fhs}, \cite{pdg}, and,
up to 2009, \cite{ngen4rev}, and some recent papers include
\cite{ngen4recent}. Most of this work has been done within the context of a
Higgs mechanism as the origin of electroweak symmetry breaking.  It has been
noted, in particular, that if the fermions of the fourth generation are
sufficiently heavy, i.e., the corresponding Yukawa couplings are sufficiently
large, then the Higgs interactions become nonperturbative and can produce
fermion condensates.  There are triviality upper limits on Yukawa couplings
obtained from fully nonperturbative dynamical-fermion lattice simulations
\cite{y}. Current experimental lower limits on the masses of possible
sequential fourth-generation quarks and leptons are given in \cite{pdg} but
depend on various assumptions such as the ordering of masses of the
fourth-generation quarks and leptons and the values of relevant mixing angles.

  Before proceeding, some remarks on the current status of (extended)
technicolor models are in order.  These theories are very ambitious, since they
incorporate a dynamical origin not just for electroweak symmetry breaking, but
also for fermion masses.  This contrasts with the standard model, which obtains
electroweak symmetry breaking by fiat, from the choice of the sign of the
coefficient of the quadratic term in the Higgs potential, and accomodates, but
does not explain, the observed quark and charged lepton masses by appropriate
choices of Yukawa couplings.  (Here and below, by the term ``standard model'',
we implicitly mean an appropriate extension of the original standard model to
account for nonzero neutrino masses.)  Moreover, theories with dynamical EWSB
are subject to a number of constraints from data on flavor-changing
neutral-current processes, splitting of the $m_t$ and $m_b$ masses, precision
electroweak measurements, limits on pseudo-Nambu-Goldstone bosons, etc.  A
fully realistic theory of this type would answer such longstanding questions as
why mass ratios like $m_e/m_\mu$ have the values that they do. Given such
ambitious goals, it is perhaps not surprising that no fully realistic TC/ETC
model has yet been constructed.

  However, there have been a number of important advances in this area.  It was
shown that technicolor theories can exhibit a large but slowly running
(``walking'') coupling \cite{holdom}, as a consequence of an approximate
infrared fixed point in the renormalization group equation for the TC gauge
coupling \cite{wtc1}-\cite{rsc97}.  Recently, there has been important progress
in elucidating the properties of walking gauge theories by means of lattice
simulations. For the case considered here, of (techni)fermions in the
fundamental representation, these recent lattice studies include the works in
Ref. \cite{lgt}. There have also been studies of walking gauge theories with
(techni)fermions in higher-dimensional representations (a few include
\cite{higherrep1,higherrep2}; see also the review \cite{sanrev}).  The walking
behavior enables TC/ETC theories to generate sufficiently large fermion masses
to match experiment with ETC mass scales that are large enough to avoid
excessive flavor-changing neutral-current effects.  Furthermore, this walking
behavior may be able to reduce technicolor contributions to $W$ and $Z$
propagators to a level in agreement with experimental limits (e.g.,
\cite{scalc}-\cite{sgf}, \cite{dewsb,sanrev}, and references therein).  One of
the results obtained from a detailed study of a reasonably ultraviolet ETC
theory was the demonstration \cite{nt,ckm} that models of this type, in which
the SM fermions transform as vectorial representations of the ETC gauge group,
did not have as severe problems with flavor-changing neutral-current processes
as had previously been thought on the basis of less ultraviolet-complete
models.  For example, one of the most severe constraints had been considered to
arise from $K^0 - \bar K^0$ mixing.  However, as was pointed out in \cite{ckm},
this would proceed via the $d \bar s$ in the $K^0$ producing a virtual $V^1_2$
ETC gauge boson (where the numbers are the gauged generational indices), but
the $s \bar d$ in the final-state $\bar K^0$ can only be produced by a
$V^2_1$. Hence, the $K^0 - \bar K^0$ transition can only proceed via a
nondiagonal ETC gauge boson mixing, $V^1_2 \to V^2_1$.  Having an
ultraviolet-complete ETC theory, one could calculate this mixing
quantitatively; this was done in Refs. \cite{nt,ckm}, and it was found to
suppress the $K^0 - \bar K^0$ transition strongly. 

\hspace{-1in}

The central role that the number of standard-model fermion generations,
$N_{gen.}$, plays in the structure of TC/ETC theories may be contrasted with
the rather different role that it plays in the standard model, supersymmetric
extensions thereof, and (supersymmetric) grand unified theories.  In these
three types of theories, at least when viewed as pointlike field theories
without being derived from a string theory, one puts in the value of $N_{gen.}$
as copies of the fermion representations. The number $N_{gen.}=3$, as such,
does not play a direct role in the symmetry breaking of the grand unified
symmetry or in the determination of the fermion masses.  Besides the (extended)
technicolor approach, only very few other approaches have attempted to predict
$N_{gen.}$ from intrinsic properties of the model rather than inserting it by
hand.  One effort in this direction was based on composite models of SM
fermions \cite{marshak83}.  A particularly elegant approach is provided by
string theory, in which $N_{gen.}$ is determined by the topology of the
compactification manifold or orbifold $M$ (and resultant number of zero modes
of the Dirac operator), as given by $|\chi(M)|/2$, where $\chi$ is the Euler
characteristic of $M$ \cite{gsw}.  Here we focus on the more bottom-up approach
provided by extended technicolor.

\section{General Theoretical Framework and Calculational Methods} 
\label{methods}

\subsection{Gauge Group} 

We consider a ($(3+1)$-dimensional) gauge theory with the gauge group
\beq
G = {\rm SU}(N_{ETC})_{ETC} \times {\rm SU}(N_{HC})_{HC} \times G_{SM} \ , 
\label{g}
\eeq
where
\beq
G_{SM} = {\rm SU}(3)_c \times {\rm SU}(2)_L \times {\rm U}(1)_Y
\label{gsm}
\eeq
is the standard-model gauge group and ${\rm SU}(N_{ETC})_{ETC}$ is the extended
technicolor gauge group, which dynamically breaks to the technicolor group
${\rm SU}(N_{TC})_{TC}$ in a series of stages at successively lower and lower
energy scales.  In order to communicate the electroweak symmetry breaking to
the standard-model fermions, the ETC group gauges the generational indices and
combines them with the technicolor indices, so that
\beq
N_{ETC} = N_{gen.} + N_{TC} \ . 
\label{netc}
\eeq
As in Refs. \cite{nt,ckm}, we take $N_{TC}=2$ because this minimizes
technicolor corrections to $W$ and $Z$ propagators.  As we will demonstrate, it
also allows us to obtain walking behavior for the residual technicolor sector
for each of the values of $N_{gen.}$ that we study, generalizing the previous
success in obtaining walking behavior for $N_{gen.}=3$. The extended
technicolor sector is arranged to be an asymptotically free chiral gauge
theory.  In earlier ETC models \cite{at94}, \cite{nt}-\cite{nag06} with
$N_{gen.}=3$, in order to obtain the desired sequential breaking of the ETC
gauge symmetry, one included another strongly coupled gauge interaction, called
hypercolor (HC). The hypercolor gauge group was taken to be ${\rm SU}(2)_{HC}$.
There were several reasons for this choice, including minimality and the fact
that SU(2) has only (pseudo)real representations, so that there are no gauge
anomalies; this gives one added flexibility in choosing the representations of
hypercolored fermions.  As noted above, one of the questions that we will
examine here is whether for the lower values of $N_{gen.}=1$ or 2, it might be
possible to simplify the model by eliminating the hypercolor interaction, so
that all of the ETC symmetry breaking is produced by ETC itself, as
self-breaking.  To anticipate our results, and as indicated in Eq. (\ref{g}),
the models that we have constructed with the requisite ETC breaking patterns
still rely on hypercolor.  We take the gauge symmetry (\ref{g}) as our starting
point but mention that there have also been studies of ideas for deriving
$N_{gen.}$ from a higher unification of gauge symmetries in a TC/ETC context
\cite{fs}.

\subsection{Fermion Content}

The fermion content of each the models includes $N_{gen.}$ generations of
standard-model quarks and leptons, arranged together with technifermions with
the same SM quantum numbers in the following ETC multiplets:
\begin{widetext} 
\beq
Q_L: \ (N_{ETC},1,3,2)_{1/3,L} \ , \quad\quad 
u_R: \ (N_{ETC},1,3,1)_{4/3,R} \ , \quad\quad 
d_R: \ (N_{ETC},1,3,1)_{-2/3,R} \ ,
\label{quarks}
\eeq
and
\beq
L_L: \ (N_{ETC},1,1,2)_{-1,L}, \quad\quad  
e_R: \ (N_{ETC},1,1,1)_{-2,R} \ . 
\label{leptons} 
\eeq
\end{widetext}
Here the numbers in parentheses refer to the dimensions of the representations
under ${\rm SU}(N_{ETC})_{ETC} \times {\rm SU}(2)_{HC} \times {\rm SU}(3)_c
\times {\rm SU}(2)_L$ and the subscript gives the weak hypercharge, $Y$. We
will also use the notation $U$, $D$, $E$, and $n$ for the
technifermions with the indicated SM quantum numbers.  Thus, for example, for
the case $N_{gen.}=3$,
\beq
e_R \equiv (e^1,e^2,e^3,e^4,e^5)_R \equiv (e,\mu,\tau,E^4,E^5)_R \ . 
\label{erexample}
\eeq
Note that the set (\ref{leptons}) does not contain right-handed,
electroweak-singlet neutrinos; these will arise as residual components of
SM-singlet fermions that transform according to larger representations of the
ETC gauge group. Refs. \cite{ckm,kt} also investigated a different set of
fermion representations with $Q_L: \ (5,1,3,2)_{1/3,L}$, $u_R: \
(5,1,3,1)_{4/3,R}$, $d_R: \ (\bar 5,1,3,1)_{-2/3,R}$, $L_L: \ (\bar
5,1,1,1)_{-1,L}$, and $e_R: \ (5,1,1,1)_{-2,R}$, and containing a corresponding
set of SM-singlets that rendered the theory anomaly-free. However, while that
set of fermions produces a natural splitting in $m_t$ and $m_b$ (and $m_c$ and
$m_s$) without excessive violation of custodial symmetry, it leads to
flavor-changing neutral current effects \cite{ckm,kt,dml} that are too large.
Other TC/ETC strategies for getting large splitting between $m_t$ and $m_b$
include the use of more than one ETC group (e.g., \cite{met} and references
therein) and the use of new gauge interaction(s) such as topcolor
\cite{topcolor}. We do not pursue these here, although it could be of interest
in future work to study such models with variable $N_{gen.}$.

Masses for quarks and leptons arise from diagrams in which these fermions emit
virtual ETC gauge bosons, making transitions to virtual technifermions, and
then reabsorb the ETC gauge bosons.  Since the contributions of the
corresponding Feynman integrals to these masses depend on the ETC gauge boson
masses via their propagators, exchanges of the heaviest ETC gauge bosons, with
masses of order the highest scale of ETC symmetry breaking, produce the
smallest fermion masses, namely those of the first generation.  For this
reason, this highest ETC breaking scale is denoted $\Lambda_1$.  In
quasi-realistic ETC models with $N_{gen.}=3$, exchanges of ETC gauge bosons
with masses of order the next lower ETC symmetry breaking scale play a dominant
role in determining the masses of second-generation SM fermions, motivating the
notation $\Lambda_2$ for this scale, and similarly for the lowest ETC scale,
$\Lambda_3$, and the third generation. These exchanges give rise to the
diagonal elements of the respective $3 \times 3$ mass matrices of the quarks of
charge 2/3 and $-1/3$ and the charged leptons.  Incorporating ETC gauge boson
mixing is required to produce off-diagonal elements of these mass matrices, as
studied in \cite{ckm,kt}. The observed quark mixing matrix arises from the
differences in the mixings in the charge 2/3 and $-1/3$ quark sectors.
Accounting for the very light neutrino masses requires an additional mechanism
in which ETC gauge boson mixing yields Dirac neutrino mass terms connecting
left- and right-handed neutrinos, and Majorana masses for right-handed
(electroweak-singlet) neutrinos, leading to a low-scale seesaw
\cite{nt,lrs,ckm}.  In quasi-realistic $N_{gen.}=3$ ETC models that are
reasonably ultraviolet-complete (at least up to $10^4$ TeV), such as those
studied in Ref. \cite{nt,ckm}, one typically has $\Lambda_1 \sim 10^3$, TeV,
$\Lambda_2 \sim 10^2$ TeV, and $\Lambda_3 \sim$ few TeV. These values are
sufficient to produce the requisite fermion masses and also to satisfy
constraints from flavor-changing neutral-current processes.

\subsection{Sequential ETC Symmetry Breaking} 

Generalizing the process of sequential ETC gauge symmetry breaking to the case
of present interest, in which $N_{gen.}$ is a variable rather than being fixed
at the physical value of 3, we require that the model have the property that 
the ETC symmetry breaking occurs in $N_{gen.}$ sequential stages,
\beq
{\rm SU}(N_{ETC})_{ETC} \to {\rm SU}(N_{ETC}-1)_{ETC} \quad {\rm at} 
\ \ \Lambda_1 \ , 
\label{lam1break}
\eeq
then, for $N_{gen.} \ge 2$, 
\beq
{\rm SU}(N_{ETC}-1)_{ETC} \to {\rm SU}(N_{ETC}-2)_{ETC} \quad {\rm at} 
\ \ \Lambda_2 \ , 
\label{lam2break}
\eeq
and so forth, down to 
\beq
{\rm SU}(3)_{ETC} \to {\rm SU}(2)_{TC} \quad {\rm at} \ \ \Lambda_{N_{gen.}} \
, 
\label{su3tosu2}
\eeq
with
\beq
\Lambda_1 > \Lambda_2 > ... > \Lambda_{N_{gen.}} \ .
\label{breakingpattern}
\eeq
For a given value of $N_{gen.}$, it is thus necessary to choose the fermion
content so that the ETC theory undergoes this requisite sequential gauge
symmetry breaking.  If one is trying to construct a quasi-realistic TC/ETC
model, then one has to do more than just obtaining the sequential ETC symmetry
breaking at the scales in Eq.  (\ref{breakingpattern}); one also has to ensure
that the actual values of $\Lambda_j$ with $j=1,2,3$ yield acceptable SM
fermion masses and mixings that are in reasonable agreement with experimental
values. However, because of the strongly coupled nature of the physics, it is
difficult to calculate the scales $\Lambda_i$ precisely.  To avoid electroweak
symmetry breaking at too high a scale, the fermions that are responsible for
this ETC symmetry breaking are taken to SM-singlets. Because the full theory is
chiral, there are no fermion mass terms in the Lagrangian.  The chiral fermions
transform according to a set of representations $\{R_i\}$ of the ETC and HC
groups.  We denote the running gauge couplings of these groups at the reference
energy scale $E=\mu$ as $g_{_{ETC}} \equiv g_{_{ETC}}(\mu)$ and $g_{_{HC}}
\equiv g_{_{HC}}(\mu)$, and we define $\alpha_{_{ETC}}=g_{_{ETC}}^2/(4\pi)$
and $\alpha_{_{HC}}=g_{_{HC}}^2/(4\pi)$.  (The implicit $\mu$-dependence of
these couplings will generally be suppressed in the notation.)  The fermion
content of the ETC and HC theories is chosen to incorporate the property that
these two interactions are both asymptotically free.  Hence, as the reference
energy scale $\mu$ decreases from high values, $\alpha_{_{ETC}}$ and
$\alpha_{_{HC}}$ increase.  As the scale $\mu$ decreases through
$\mu=\Lambda_1$, the HC and ETC interactions produce bilinear fermion
condensates.  Because of the chiral nature of the ETC fermion representations,
these condensates generically break the ETC gauge symmetry.  The fermions
involved in the condensates gain dynamical masses of order $\Lambda_1$ and the
gauge bosons corresponding to broken generators gain masses of order
$g_{_{ETC}}\Lambda_1$.

  Following the principles of effective field theory, one analyzes the
evolution of the theory to lower energy scales by integrating out these fields
that gain masses at the scale $\Lambda_1$.  One then proceeds to study the
evolution of the low-energy effective field theory at energy scales $\mu <
\Lambda_1$.  For $N_{gen.}=1$, this is the only stage of ETC symmetry breaking,
while for $N_{gen.} \ge 2$, one analyzes the evolution of each of the
$N_{gen.}$ effective field theories resulting from the corresponding stages of
symmetry breaking and resultant acquisition of masses by the fermions involved
in the condensates and by the gauge bosons corresponding to broken generators
at each stage.  The overarching ETC theory considered here is designed so that
at each level of symmetry breaking, the descendent ETC and HC interactions
remain asymptotically free, and hence the respective ETC and HC gauge couplings
continue to grow, triggering the formation of the next set of condensates and
the next stage of symmetry breaking \cite{lambhc}.

\subsection{Resultant Technicolor Theory} 

  The result of the $N_{gen.}$ stages of ETC symmetry breaking is a theory
invariant under the gauge group 
\beq
{\rm SU}(2)_{TC} \times {\rm SU}(2)_{HC} \times G_{SM} \ . 
\label{gtcghcgsm}
\eeq
Since the only HC-nonsinglet fermions are SM-singlets, their condensates are
automatically invariant under $G_{SM}$. In contrast, the model is designed so
that the technifermion condensates transform as operators with weak isospin $T
=1/2$ and weak hypercharge $|Y|=1$, so these break the electroweak group 
${\rm SU}(2)_L \times {\rm U}(1)_Y$ to electromagnetic U(1)$_{em}$.
On the basis of vacuum alignment arguments, one infers that the technifermion
condensates include 
\beq
\langle \bar F_{i,L} F^i_R \rangle + h.c. \ , 
\label{technicondensate}
\eeq
where here $F$ refers to $U$, $D$, and $E$ and the sum over $i$ is over
SU(2)$_{TC}$ gauge indices.  For technineutrinos, the left- and right-handed
components in the models of Refs. \cite{nt,ckm} and also in most of our present
models transform according to conjugate representations of SU(2)$_{TC}$, and
hence the condensates for these are $\langle \epsilon^{ij} \bar n_{i,L} n_{j,R}
\rangle + h.c.$, leading to the effective mass terms 
\beq
-\Lambda_{TC} \, \sum_i \, \epsilon^{ij} \, \bar n_{i,L} n_{j,R} + h.c. 
\label{mnu_dirac_tc}
\eeq
(In the specific models below, the $n_{j,R}$ will arise from various ETC
representations with different labels, such as $\psi_{j,R}$; we use the
$n_{j,R}$ notation here to indicate the technineutrino components.)

The $W$ and $Z$ bosons gain masses given by 
\beq 
m_W^2 = \frac{g^2 F_{TC}^2 N_D}{4}
\label{mwsq}
\eeq
and
\beq
m_Z^2 = \frac{(g^2+g'^2) F_{TC}^2 N_D}{4} \ , 
\label{mzsq}
\eeq
where $g$ and $g'$ are the SU(2)$_L$ and U(1)$_Y$ gauge couplings and 
$N_D$ denotes the number of SU(2)$_{TC}$ technidoublets, 
\beq
N_D=N_c+1=4 \ , 
\label{ndtc}
\eeq
and $F_{TC}$ is the TC analogue of $f_\pi$. To fit experiment, $F_{TC} = 250$
GeV.

\subsection{Gauge Coupling Evolution and Criteria for Fermion Condensation} 

The evolution of the various gauge couplings is determined by the respective
beta functions.  For a given gauge group $G$ with gauge coupling $g_{_{G}}$,
the beta function is $\beta = dg_{_{G}}/dt$, where $dt = d\ln \mu$.  For our
analysis, the beta functions for the ETC and HC groups will be of particular
interest, since these are the ones that are relevant for the sequential
breaking of the ETC symmetry down to technicolor.  We have 
\beq
\beta(g_{_{G}}) = -g_{_{G}} \, \sum_{\ell=1}^\infty \, b_{G,\ell} ,
\left ( \frac{g_{_{G}}^2}{16 \pi^2} \right )^\ell 
\label{beta}
\eeq
where $b_{G,\ell}$ arises at $\ell$-loop order in perturbation theory, and we
will focus on the first two coefficients, $b_{G,1}$ and $b_{G,2}$, since they
are the only scheme-independent ones. Equivalently, with
$\alpha_{_{G}}=g_{_{G}}^2/(4\pi)$,
\beq
\frac{d\alpha_{_{G}}}{dt} = - \frac{\alpha_{_{G}}^2}{2\pi}
\left [ b_{G,1} + \frac{b_{G,2} \, \alpha_{_{G}}}{4\pi} + 
O(\alpha_{_{G}}^2) \right ] \ . 
\label{betaalpha}
\eeq
We will apply these results for $G$ equal to the ETC, TC, and HC groups,
respectively.  To avoid cumbersome notation, henceforth we will suppress the
subscript $G$ where no confusion will result.  Since the ETC and HC
interactions are asymptotically free, it follows that in each case, $b_1 > 0$.
If there are sufficiently many fermions that are nonsinglets under a given
interaction, $b_2$ reverses sign from positive to negative, and, in this case,
the perturbative beta function has a infrared zero away from the origin, given,
to this order, by
\beq
\alpha_{IR} = - \frac{4\pi b_1}{b_2} \ . 
\label{alfir}
\eeq
This perturbative IR zero is important as a natural origin for walking
technicolor.  This will be especially important for the one-family technicolor
sector incorporated in our models, since an analysis using the two-loop beta
function and the Dyson-Schwinger equation for the technifermion propagator
suggests that this theory exhibits an approximate infrared zero given by
Eq. (\ref{alfir}) and resultant walking behavior \cite{wtc1,wtc2}.  (We note
that a beta function may also exhibit a nonperturbative infrared zero away from
the origin \cite{brodskyfreeze,lmax,creutz}.) 

Let us, then, consider a possible channel for chiral fermions, 
transforming as representations $R_1$ and $R_2$ under a given gauge group, to
form a bilinear condensate transforming as $R_{cond.}$:
\beq
R_1 \times R_2 \to R_{cond.} \ . 
\label{channel}
\eeq
An approximate measure of the attractiveness of this condensation channel is
\beq
\Delta C_2 = C_2(R_1) +C_2(R_2) - C_2(R_{cond.}) \ , 
\label{deltac2}
\eeq
where $C_2(R)$ is the quadratic Casimir invariant for the representation $R$
\cite{casimir}.  A solution of the Dyson-Schwinger equation for a fermion
propagator with zero input mass, in the approximation of single gauge boson
exchange, yields a solution with a nonzero, dynamically generated mass if the
gauge coupling exceeds the critical value $\alpha_{cr}$ given by
\beq
\frac{3 \, \Delta C_2 \, \alpha_{cr}}{2 \pi} = 1 \ . 
\label{alfcrit}
\eeq
Corrections to this estimate have been studied in Ref. \cite{alm}, but it will
be sufficient for our purposes here. Since the dynamically generated mass
multiplies the corresponding bilinear operator for this fermion in the
effective Lagrangian, this is equivalent to the formation of a condensate for
this operator.  Some general formulas that are used in our calulations of the
beta functions for the models are given in the appendix. 

We comment on some other necessary conditions that an acceptable ETC model
should meet. First, a model needs to satisfy the condition that the dynamical
gauge symmetry breaking occurs at the highest scale in such a manner as not to
break electroweak symmetry.  This is not guaranteed, since, given that the
SM-nonsinglet fermions (and technifermions) transform as specified in
Eqs. (\ref{quarks}) and (\ref{leptons}), the theory 
contains the highly attractive possible condensation channel
\beq
N_{_{ETC}} \times \bar N_{_{ETC}} \to 1 \ , \ {\rm with} \ \Delta C_2 = 
\frac{N_{_{ETC}}^2-1}{N_{_{ETC}}}  
\label{ffbar}
\eeq
involving these SM-nonsinglet fermions.  This condensation channel must be
avoided in the ETC theory for a number of reasons: (i) it would break
electroweak symmetry at too high a scale; (ii) it would not break the ETC gauge
symmetry, and hence (iii) it would not separate the usual SM fermions from the
technifermions, and could produce a model in which all of these fermions are
confined by the ETC interaction.  In quasi-realistic models (which have
$N_{gen.}=3$) such as those of Refs. \cite{at94,nt,ckm,nag06}, one avoids the
occurrence of the unwanted condensation in the channel (\ref{ffbar}) by a
hybrid mechanism which makes use of the fact that the model contains SM-singlet
chiral fermions transforming as higher-dimensional representations of the ETC
gauge group, and these can condense in a first stage of ETC symmetry breaking.
For the second and third stages of ETC symmetry breaking, the models of
Refs. \cite{at94,nt,ckm,nag06} rely on the effects of the auxiliary strongly
coupled hypercolor gauge interaction. Another comment pertains to the
sequential breaking of the ETC gauge symmetry.  The model should be constructed
in a manner that although this symmetry is chiral, the residual TC gauge
symmetry is vectorial and unbroken, so that it confines and produces the
necessary bilinear technifermion condensates.  The one-family technicolor
sectors in the models that we consider here satisfy this condition.

\section{A Model With $N_{gen.}=1$}
\label{model_ngen1}

\subsection{Field Content} 

We first study an ETC model that contains a single generation of standard-model
fermions, $N_{gen.}=1$. From Eq. (\ref{netc}), with the choice $N_{TC}=2$, it
follows that $N_{ETC}=3$, so that the ETC gauge group is SU(3)$_{ETC}$. One
constructs the ETC model to have a single stage of gauge symmetry breaking,
viz.,
\beq
{\rm SU}(3)_{ETC} \to {\rm SU}(2)_{TC} \ . 
\label{su32}
\eeq
The fermions which are nonsinglets under the SM gauge group are as given in
Eqs. (\ref{quarks}) and (\ref{leptons}) with $N_{ETC}=2$. We take the
SM-singlet fermions to be
\beq
\psi_{i,R} \ : \quad (\bar 3,1,1,1)_{0,R}
\label{ngen1_psi}
\eeq
\beq
\chi^{i,\alpha}_R \ : \quad (3,2,1,1)_{0,R}
\label{ngen1_chi}
\eeq
and
\beq
\omega_{\alpha,R} \ : \quad (1,2,1,1)_{0,R} \ , 
\label{ngen1_omega}
\eeq
where the numbers refer to the representations of ${\rm SU}(3)_{ETC} \times
{\rm SU}(2)_{HC} \times {\rm SU}(3)_c \times {\rm SU}(2)_L$ and the subscripts
denote the weak hypercharge, $Y$. This is an anomaly-free chiral gauge theory.
Here we define the SU(3)$_{ETC}$ index $i=1$ to refer to the single SM
generation, while the indices $i=2,3$ are SU(2)$_{TC}$ indices. Note that since
the total number of chiral fermions transforming according to the fundamental
representation of SU(2)$_{HC}$ is even (equal to four), the theory is free of a
global Witten anomaly associated with the homotopy group $\pi_4({\rm
SU}(2))={\mathbb Z}_2$.

\subsection{Condensation at $\Lambda_1$ Breaking SU(3)$_{ETC}$ to SU(2)$_{TC}$}

We proceed to study the evolution of this theory from high energy scales.  The
SU(3)$_{ETC}$ gauge interaction is asymptotically free, and the first two
coefficients of the ETC beta function are $b_1 = 5$ and $b_2 = -12$.  By
Eq. (\ref{alfir}), the perturbative two-loop beta function for the
SU(3)$_{ETC}$ theory thus has a zero at $\alpha_{_{ETC,IR}} = 5\pi/3$.
Obviously, this prediction has considerable theoretical uncertainty because of
the large value of the coupling.

The SU(2)$_{HC}$ sector is asymptotically free, with fermion content consisting
of four chiral fermions, or equivalently, two Dirac fermions, transforming
according as hypercolor doublets. From Eqs.  (\ref{b1hc}) and (\ref{b2hc}) with
$N_{f,1/2}=2N_{f,D,1/2}=4$, it follows that the first two coefficients of the
SU(2)$_{HC}$ beta function are $b_1 = 6$ and $b_2 = 29$.  Since $b_2$ has the
same sign as $b_1$, this perturbative HC beta function does not have a zero
away from the origin.  The value $N_{f,D,1/2}=2$ is well below the critical
value, $N_{f,cr} \simeq 8$, where, according to the analysis of the
Dyson-Schwinger equation for the fermion propagator (discussed further in the
appendix), the theory would go over from one with confinement and spontaneous
chiral symmetry breaking (S$\chi$SB) to a chirally symmetric one, which is
often plausibly inferred to be a non-Abelian Coulomb phase.  This property is
important for our use of the HC interaction in this model, since it implies
that as the energy scale $\mu$ decreases, the HC interaction will produce
condensates of HC-nonsinglet fermions.

Since this theory has two asymptotically free gauge interactions, SU(3)$_{ETC}$
and SU(2)$_{HC}$, the properties of the theory depend on the relative sizes of
the running couplings $\alpha_{_{ETC}}$ and $\alpha_{_{HC}}$ at a given
reference energy scale $\mu$.  In order to obtain the desired pattern of
symmetry breaking, we choose the initial value of the HC coupling at a high
scale to be sufficiently large that the HC interaction plays the dominant role
in breaking SU(3)$_{ETC}$ to SU(2)$_{TC}$.  This strategy is used because the
the ETC interaction, by itself, would favor condensation in the undesired
channel (\ref{ffbar}), viz., in this case $3 \times \bar 3 \to 1$, involving
the left and right-handed SM-nonsinglet fermions and technifermions.  This
unwanted condensation is avoided by making the HC interaction strong enough to
play the controlling role in determining the formation of fermion condensates.

Thus, as the energy scale $\mu$ decreases from high values through a value that
we shall denote as $\Lambda_1$, the HC and ETC gauge couplings grows
sufficiently large that together they produce a bilinear fermion condensate.
Given the dominant role of the HC interaction, the most attractive channel
involves the condensation of the $\chi^{i,\alpha}_R$. This channel is
\beq
(3,2,1,1)_0 \times (3,2,1,1)_0 \to (\bar 3,1,1,1)_0
\label{su3etcmac}
\eeq
with $\Delta C_2 = 4/3$ for SU(3)$_{ETC}$ and $\Delta C_2 = 3/2$ for
SU(2)$_{HC}$. The associated condensate is
\beq
\langle \epsilon_{\alpha \beta} \epsilon_{ijk} \chi^{j \alpha \ T}_R C 
\chi^{k \beta}_R \rangle \ , 
\label{su3etc_chichicondensate}
\eeq
where $\epsilon_{ijk}$ and $\epsilon_{\alpha\beta}$ are the totally
antisymmetric tensor densities for SU(3)$_{ETC}$ and SU(2)$_{HC}$,
respectively.  (Condensation of the hermitian conjugate operator is implicitly
understood to occur here and below.)  As noted, the HC attraction plays a
crucial role here, since as far as the ETC interaction itself is concerned, the
channel (\ref{su3etcmac}) is less attractive than the channel $3 \bar \times 3
\to 1$, with $\Delta C_2 = 2C_2(\Box) = 8/3$, involving the left and right
chiral components of the SM fermions, which are HC-singlets. 
The condensation (\ref{su3etcmac}) breaks
${\rm SU}(3)_{ETC}$ to ${\rm SU}(2)_{TC}$ and is invariant under SU(2)$_{HC}$.
With no loss of generality, we choose the uncontracted SU(3)$_{ETC}$ index to
be $i=1$, so that, carrying out the sum over repeated SU(3)$_{ETC}$ indices in
Eq. (\ref{su3etc_chichicondensate}), one has, for the actual condensate,
\beq
2\langle \epsilon_{\alpha \beta} \chi^{2, \alpha \ T}_R C 
\chi^{3 \beta}_R \rangle \ . 
\label{su3etc_chichicondensate_explicit}
\eeq
This condensate involves only SM-singlet fermions. With this symmetry breaking,
the $i=1$ components of the various fermion multiplets with nonsinglet SM
quantum numbers split off from the other two components to become the first
generation of SM fermions, while the remaining components, with $i=2,3$, are
SU(2)$_{TC}$ technifermions.  The fermions $\chi^{j \alpha \ T}_R$ with $j=2,3$
involved in the condensate gain dynamical masses of order $\Lambda_1$, and the
five gauge bosons in the coset space ${\rm SU}(3)_{ETC}/{\rm SU}(2)_{TC}$ gain
masses of order $g_{_{ETC}} \Lambda_1 \simeq \Lambda_1$.

\subsection{Condensation at $\Lambda_1' < \Lambda_1$}

The effective theory operative at energy scales just below $\Lambda_1$ is
invariant under the strongly coupled gauge symmetries ${\rm SU}(2)_{TC} \times
{\rm SU}(2)_{HC}$. Given that the HC interaction plays a dominant role in the
formation of condensates, another one that forms in this theory at a scale that
we shall denote $\Lambda_1'$, slightly below $\Lambda_1$, is driven by the HC
interaction alone, without the help of SU(2)$_{TC}$.  This involves the
remaining component of the original $\chi^{i,\alpha}_R$ field, namely, 
$\chi^{1,\alpha}_R$, together with $\omega_{\alpha,R}$.  These plausibly
condense via the channel
\beq
(1,2,1,1)_0 \times (1,2,1,1)_0 \to (1,1,1,1)_0 \ , 
\label{chi1omegacondensate}
\eeq
where here the first number in the parentheses refers to the dimensionality of
the representations of these fields under SU(2)$_{TC}$.  The reason that this
occurs at a scale somewhat below $\Lambda_1$ is that the channel
(\ref{chi1omegacondensate}) has the same measure of attractiveness as regards
the HC interaction as the most attractive channel (\ref{su3etcmac}), namely,
$\Delta C_2 = 3/2$, but does not receive any additional attraction from the TC
interaction.  The associated condensate is
\beq
\langle \chi^{1,\alpha \ T}_R C \omega_{\alpha,R} \rangle \ . 
\label{chi1wcondensate}
\eeq
This respects the same residual symmetry group as the condensate
(\ref{su3etc_chichicondensate_explicit}). As a result of the formation of this
condensate, the $\chi^{1,\alpha}_R$ and $\omega_{\alpha,R}$ fermions gain
dynamical masses of order $\Lambda_1'$. In QCD the ratio of the lowest-lying
($J^{PC}=0^{++}$) glueball mass to the QCD scale $\Lambda_{QCD} \simeq 250$ MeV
is about $\kappa_g \simeq 7$.  The present model (and the models with higher
values of $N_{gen.}$) would also contain SU(2)$_{HC}$-singlet bound states of
hypercolor gluons at scales of order $\kappa_g \Lambda_1$ and
SU(2)$_{TC}$-singlet bound states of technigluons at scales of order $\kappa_g
\Lambda_{TC}$.  The actual mass eigenstates would be comprised of
linear combinations of purely gluonic and fermion-antifermion states.

\begin{figure}[hbtp]
\begin{center}
\includegraphics[3in, 8.7in][4.5in, 10in]{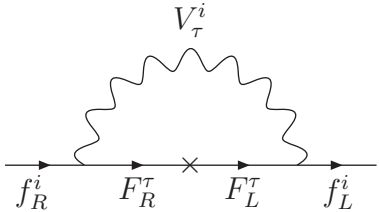}
\end{center}
\caption{\footnotesize{Graph generating a SM fermion mass term,
$m_{f^i} \bar f_{i,L} f^j_R$, diagonal in generational indices $i$. Here
$f$ stands for a SM quark or charged lepton, $F$ for the corresponding
technifermion, and $\tau$ for an SU(2)$_{TC}$ index.}}
\label{ffdiag}
\end{figure}

\subsection{Theory at Energy Scales Below $\Lambda_1'$ and SM Fermion Mass 
Generation} 

The masses of quarks and changed leptons arise from the one-loop diagram shown
in Fig. 1. Here and below, it is understood that higher-loop diagrams also make
important contributions because of the strong technicolor dynamics.  The
resultant value of the fermion masses for this $N_{gen.}=1$ case is
\beq
m_f \simeq \kappa \, \eta \, \frac{\Lambda_{TC}^3}{\Lambda_1^2} \ , 
\label{mf_gen1}
\eeq
where $\kappa$ is a numerical factor of O(10) (computed in Ref.
\cite{ckm}), and $\eta$ is the renormalization factor, reflecting the shift in
the running mass between the scales $\Lambda_{TC}$ and $\Lambda_1$:
\beq
\eta = \exp \left [ \int_{\Lambda_{TC}}^{\Lambda_1} \frac{d\mu}{\mu} \, 
\gamma(\alpha(\mu)) \right ] \ , 
\label{eta}
\eeq
where $\gamma$ is the anomalous dimension for the technifermion bilinear and
$\alpha$ refers to the TC coupling, as inherited from the enveloping ETC
theory. With $N_{f,cr} \simeq 8$, as indicated by the beta function and
Dyson-Schwinger analysis \cite{wtc2}, the TC theory exhibits the slowest
evolution of the coupling between $\Lambda_{TC}$ and $\Lambda_1'$, where there
are 8 Dirac technifermions active.  Since the anomalous dimension for the
technifermion bilinear is generically $\gamma \simeq O(1)$ near the
approximate IR fixed point of the renormalization group equation for the TC
interaction, it follows that $\eta \sim \Lambda_1'/\Lambda_{TC}$, so that
\beq
m_f \simeq \kappa \, \frac{\Lambda_1' \Lambda_{TC}^2}{\Lambda_1^2} 
 \quad {\rm for \ walking \ up \ to} \ \ \Lambda_1' \ . 
\label{mf_gen1_walktolamp}
\eeq
In addition, owing to the presence of the approximate IR fixed point, the TC
theory presumably also exhibits at least some walking behavior in the
higher-lying interval between $\Lambda_1'$ and $\Lambda_1$, where there are 9
Dirac technifermions active (including the $\psi_{j,R}$ with $j=2,3$).  To the
extent that the theory has walking behavior all the way up to $\Lambda_1$ with
an associated large anomalous dimension for the technifermion bilinear, the
renormalization factor would be larger, viz., $\eta \sim
\Lambda_1/\Lambda_{TC}$, and hence 
\beq
m_f \simeq \kappa \, \frac{\Lambda_{TC}^2}{\Lambda_1}  \quad 
{\rm for \ walking \ up \ to} \ \ \Lambda_1 \ . 
\label{mf_gen1_walking}
\eeq
In this case, with $\kappa \sim O(10)$, the model has the interesting feature
that the fermion masses are comparable to the electroweak symmetry-breaking
scale, given by $\Lambda_{TC}$.  Even in the case where the TC theory exhibits
walking only up to $\Lambda_1'$, the fermion masses are considerably larger
than the value $\kappa \Lambda_{TC}^3/\Lambda_1^2$ that they would have had in
the absence of walking, since $\Lambda_1'/\Lambda_{TC} > 1$.  The ratio
$\Lambda_1'/\Lambda_1$ depends on the relative strength of the HC and ETC
interactions, as measured by the ratio of the running couplings squared,
$\alpha_{_{HC}}/\alpha_{_{ETC}}$. The larger this ratio is, the smaller is the
ratio $\Lambda_1'/\Lambda_1$.  We find that the model does not have enough
structure to lead to a low-scale seesaw mechanism for small, nonzero neutrino
masses, as presented in Ref. \cite{nt} for a $N_{gen.}=3$ model.  Further
ingredients would be necessary to produce such a seesaw.

At first sight, this model would appear to predict that for the fermions of the
single generation, the masses of the charge 2/3 quark, the charge $-1/3$ quark,
and the charged lepton are all equal.  However, here the walking behavior of
the technicolor theory can play yet another important role.  Although the SM
gauge interactions are small at the scale $\Lambda_{TC}$ where the electroweak
symmetry breaking occurs, such small perturbations would have a magnified
effect on fermion masses in a theory with walking behavior.  In turn, this has
the potential to explain the relative sizes of the up-type, down-type, and
charged lepton masses.  The fact that the SU(3)$_c$ interaction contributes an
attractive force that aids in the formation of the bilinear techniquark
condensates $\langle \bar U_L U_R\rangle + h.c.$ and $\langle \bar D_L
D_R\rangle + h.c.$ would mean that these condensates would naturally form at a
somewhat higher scale than the technilepton condensate, and hence the
dynamically generated mass for the techniquarks would be somewhat larger than
that for the technileptons.  This would provide a natural explanation for why
the quarks of a given generation (a single generation here and multiple
generations in other cases) have larger masses than the charged leptons.
Moreover, the U(1)$_Y$ interaction is attractive for the $\langle \bar U_L
U_R\rangle + h.c.$ condensate with $Y_{\bar U_L}Y_{U_R} = -4/9$, but repulsive
for the $\langle \bar D_L D_R\rangle + h.c.$ condensate, with $Y_{\bar
D_L}Y_{D_R} = 2/9$, so that the former condensate might be somewhat larger than
the latter, and similarly for the corresponding dynamical techniquark masses.
In this model, this would imply that the charge 2/3 quark is heavier than
the charge $-1/3$ quark.  (For leptons, the product $Y_{\bar L_L}Y_{e_R} = -2$
is also attractive.) However, one must remark that in a quasi-realistic model,
the splitting of the dynamical masses of the charge 2/3 and charge $-1/3$
techniquarks could produce an excessively large violation of custodial
symmetry. The explanation for the fact that $m_u < m_d$ for the first
generation requires consideration of off-diagonal elements in the up and
down-type quark mass matrices.

\subsection{A Variant of the $N_{gen.}=1$ Model} 

It is useful to discuss a variant of this model which provides an 
illustration of another problem of which one must be aware in TC/ETC
model-building.  This is the possibility that even if there are sufficiently
few technifermions and their representations are sufficiently small that the
technicolor interaction is asymptotically free, it may still happen that the
technicolor sector evolves into the infrared with non-Abelian Coulombic
behavior rather than confinement and spontaneous chiral symmetry breaking.
If the technicolor sector were to exhibit this behavior, it would
render the model untenable, since then the only breaking of electroweak
symmetry would be via QCD, at much too low a scale \cite{tc,smr}.

Thus, let us consider a modification of the $N_{gen.}=1$ model in
which we change the SM-singlet fermion content so that there is a right-handed
SM-singlet, HC-singlet fermion that transforms as a 3 rather than a $\bar 3$ of
SU(3)$_{ETC}$; consequently, after the breaking of SU(3)$_{ETC}$ to
SU(2)$_{TC}$, it can produce a mass term for the neutrino without any ETC gauge
boson mixing.  To ensure that there is no SU(3)$_{ETC}$ gauge anomaly, one must
also modify the other SM-singlet, ETC-nonsinglet fermion representations.  An
example of such a model has the SM-singlet fermion sector
\beq
\psi^i_{p,R}: \ : \quad 3(3,1,1,1)_{0,R} 
\label{ngen1_psi_rev}
\eeq
\beq
\chi_{i,\alpha,R} \ : \quad (\bar 3,2,1,1)_{0,R}
\label{ngen1_chi_rev}
\eeq
and
\beq
\omega^\alpha_R \ : \quad (1,2,1,1)_{0,R} \ , 
\label{ngen1_omega_rev}
\eeq
where $p=1,2,3$ is a copy index for the $\psi^i_{p,R}$ fields.  The ETC
symmetry breaking would again occur in the same manner as before, with obvious
switches of lower and upper indices, via the formation of the condensates
\beq
\langle \epsilon^{1jk}\epsilon^{\alpha \beta} \chi_{j,\alpha,R}^T C 
\chi_{k,\beta,R} \rangle = 2\langle \epsilon^{\alpha\beta}
\chi_{2,\alpha,R}^T C \chi_{3,\beta,R} \rangle \ , 
\label{ngen1_chichicondensate_rev}
\eeq
forming at the scale $\Lambda_1$ and
\beq
\langle \chi_{1,\alpha,R}^T C \omega^\alpha_R \rangle \ , 
\label{ngen1_chiomegacondensate_rev}
\eeq
forming at the somewhat lower scale $\Lambda_1'$.  The fermions involved in
these condensates (which include all of the HC-nonsinglet fermions) gain
dynamical masses of order the respective scales $\Lambda_1$ and $\Lambda_1'$
and are integrated out in the low-energy theory that is operative below
$\Lambda_1'$. 

The fermion representations of this model have been chosen so as to allow 
technineutrino mass terms of the form 
\beq
-\Lambda_{TC} \, \sum_i \, \bar n_{i,L} \psi^i_{p,R} + h.c., 
\label{mnu_dirac_tc_rev}
\eeq
where the sum on $i$ is over the TC indices, and resultant Dirac neutrino 
mass terms 
\beq
-m_D \, \sum_i \, \bar \nu_L \psi^1_{p,R} + h.c. \ , 
\label{mnu_dirac_rev}
\eeq
which could form without ETC gauge boson mixing.  Of course, this unsuppressed
Dirac neutrino mass term itself would be undesired without an appropriate
seesaw to yield appropriately small observable neutrino masses.  What we focus
on here is the possible problem with this model resulting from the fact that
the SU(2)$_{TC}$ sector has 18 chiral fermions, or equivalently, 9 Dirac
fermions.  This is slightly greater than the estimate $N_{f,cr}=8$ for the
critical number of Dirac fermions beyond which an SU(2) gauge theory would
evolve, in the infrared, in a non-Abelian Coulombic manner, such that the
coupling would approach an infrared fixed point and never get large enough to
produce spontaneous chiral symmetry breaking.  Although there is a theoretical
uncertainty of order $\Delta N_{f,cr} \sim 1$ in this estimate, this is a
concern.  In the present context, if, indeed, as the SU(2)$_{TC}$ theory
evolved to scales below $\Lambda_1'$, the coupling $\alpha_{_{TC}}$ did not
increase sufficiently to produce the bilinear technifermion condensates that
are necessary for electroweak symmetry breaking, then the theory would not be
acceptable even as a toy model of dynamical EWSB. The lesson from this model is
that in constructing TC/ETC models, there is a rather tight constraint on the
number of technifermions that should be present; this number should be large
enough so to yield an approximate infrared fixed point in the TC beta function
at a value $\alpha_{_{TC}}=\alpha_{IR}$ that is slightly larger than the
critical value $\alpha_{cr}$ for condensate formation, thereby giving rise to
walking behavior. However, this number of technifermions must not be so large
as to push $\alpha_{IR}$ below $\alpha_{cr}$, which would cause the technicolor
theory to evolve in the infrared as a non-Abelian Coulomb phase.

\section{An Illustrative $N_{gen.}=1$ Model with Only ETC Symmetries}
\label{failure}

In this section we deviate from the use of the high-scale gauge group (\ref{g})
and explore one of the questions posed at the beginning of the paper, namely
whether, for a different value of $N_{gen.}$ than the physical value, one might
be able to construct a TC/ETC model that could be simpler, in that the ETC
symmetry breaking would involve only the ETC interaction itself and not have to
make use of the additional strongly coupled hypercolor interaction.  It is
natural to investigate this question for the case of $N_{gen.}=1$, since for
this case one has what would appear to be an easier task to accomplish, namely
only one, rather than three stages of ETC symmetry breaking.  However, there is
a countervailing effect: because the resultant ETC gauge group is just
SU(3)$_{ETC}$, fermion representations that are nontrivial, distinct
representations for higher SU($N$) groups degenerate here.  This makes it more
difficult to use the ETC interaction by itself to obtain condensates in
channels that are more attractive than the channel in Eq. (\ref{ffbar}). Recall
that one wants condensation in the channel (\ref{ffbar}) only at the lowest,
technicolor, stage, not at higher ETC scales, since, among other things, such a
condensate would break electroweak symmetry at too high a scale. In the
following we will use interchangeably the notation $[k]_N$ for the rank-$k$
antisymmetric representation of SU($N$) and the Young tableau notation. For
example, for $N_{gen.} \ge 2$ and hence, by Eq. (\ref{netc}), $N_{ETC} \ge 4$,
the $[2]_{N_{ETC}}$ representation is a nontrivial distinct representation of
the ETC gauge group, but for $N_{gen.}=1$, it is equivalent to the conjugate
fundamental representation, $[\bar 1]_3$.  In general, for an SU($N$) group
with even $N=2k$, the $[k]_N$ representation is self-conjugate.  Hence, for
example, for $N_{gen.}=2$, whence $N_{ETC}=4$, a chiral fermion $\psi^{ij}_R$
that transforms as the $\asym = [2]_4$ representation can condense as $\langle
\epsilon_{ijk\ell} \psi^{ij \ T}_R C \psi^{k \ell}_R \rangle$.

For our illustrative $N_{gen.}=1$ model, we thus attempt to use, as the 
high-scale gauge group, ${\rm SU}(3)_{ETC} \times G_{SM}$ without
hypercolor.  We choose the SM-nonsinglet fermion content as given in Eqs. 
(\ref{quarks}) and (\ref{leptons}) (with the SU(2)$_{HC}$ entries implicitly
removed) and the following SM-singlet fermions:
\beq
\chi_{i,p,R} \ : 7(\bar 3,1,1,1)_0 
\label{ngen1_nohc_chi}
\eeq
\beq
\psi^{ij} \ : (6,1,1,1)_0 \ . 
\label{ngen1_nohc_psi}
\eeq
That is, we use a fermion $\psi^{ij}_R$ transforming as the symmetric rank-2
tensor representation of SU(3)$_{ETC}$, of dimension 6, with 7 copies of the
$\bar 3$ representation of SU(3)$_{ETC}$, indexed by $p \in \{1,..,7\}$. The
reason for the 7 copies is to cancel the anomaly of the symmetric tensor, which
is $N+4$ times that of the fundamental for an SU($N$) theory.

The first two coefficients of the beta function for this SU(3)$_{ETC}$ theory
are $b_1 = 2$ and $b_2 = -79$. This two-loop perturbative beta function has a
an infrared zero at
\beq
\alpha_{_{ETC},IR}= \frac{8\pi}{79} \simeq 0.32 \ . 
\label{ngen1_alletc_alfir}
\eeq
Thus, as the energy scale decreases from large values, $\alpha_{_{ETC}}$
increases toward this value. The most attractive channel for the formation of a
bilinear fermion condensate is
\beq
(\bar 3,1,1,1)_0 \times (6,1,1,1)_0 \to (3,1,1,1)_0 \ , 
\label{ngen1_alletc_mac1}
\eeq
with $\Delta C_2 = 10/3$.  If there were condensation in this channel, it
would involve the condensate
\beq
\langle \psi^{ij \ T}_R C \chi_{j,p,R} \rangle \ , 
\label{ngen1_alletc_cond}
\eeq
where, without loss of generality, we could pick $i=1$ and $p=1$.  However,
from Eq. (\ref{alfcrit}), the minimum value of $\alpha_{_{ETC}}$ for condensate
formation in this channel is
\beq
\alpha_{cr} = \frac{\pi}{5} \simeq 0.63 \ . 
\label{ngen1_alletc_alfcrit}
\eeq
Even taking account of the possible contributions of higher-order terms in the
beta function and the theoretical uncertainties in the analysis of the
Dyson-Schwinger equation, the value of $\alpha_{_{ETC,IR}}$ in Eq.
(\ref{ngen1_alletc_alfir}) is thus well below the value needed to trigger the
condensation in this channel.  

We conclude that as the SU(3)$_{ETC}$ theory evolves from high energy scales to
lower ones, it would probably not produce any condensates and instead would
probably maintain explicit chiral symmetry.  The infrared zero in the
SU(3)$_{ETC}$ beta function would thus be an exact infrared fixed point.  Of
course, the failure of the theory to break chiral symmetry with condensate
formation would prevent the splitting off of SM-nonsinglet, TC-singlet
components of quarks and leptons from the SU(3)$_{ETC}$ multiplets in
Eqs. (\ref{quarks}) and (\ref{leptons}) and, related to this, would prevent the
breaking of the SU(3)$_{ETC}$ symmetry to an SU(2)$_{TC}$ symmetry.  The
absence of any technifermion condensates breaking electroweak symmetry at the
usual scale of roughly 250 GeV would exclude this theory as a useful toy model
for dynamical electroweak symmetry breaking. Although we do not try to present
a no-go theorem precluding the construction of a TC/ETC theory with $N_{gen.}$
SM fermion generations that could accomplish all of the stages of ETC symmetry
breaking without the use of the auxiliary strongly coupled hypercolor gauge
symmetry, this example shows the type of difficulties that such a construction
can encounter, even for the simple case of $N_{gen.}=1$.

\section{A Model With $N_{gen.}=2$} 
\label{model_ngen2}

\subsection{Field Content} 

Returning to the framework of Eq. (\ref{g}), we next consider the case of two
SM fermion generations, $N_{gen.}=2$.  Substituting this in Eq. (\ref{netc})
with $N_{TC}=2$, we have $G_{ETC}={\rm SU}(4)_{ETC}$. In order to produce the
necessary hierarchical structure for the two generations of SM fermion masses,
we construct the model so that it undergoes a two-stage sequential breaking of
the ETC symmetry, namely
\beq
{\rm SU}(4)_{ETC} \to {\rm SU}(3)_{ETC} \quad {\rm at} \ \Lambda_1 
\label{gen2_lam1}
\eeq
followed by 
\beq
{\rm SU}(3)_{ETC} \to {\rm SU}(2)_{TC} \quad {\rm at} \ \Lambda_2 
\label{gen2_lam2}
\eeq
with $\Lambda_1 > \Lambda_2$. As before, in order to obtain this
symmetry-breaking pattern, we use an auxiliary SU(2)$_{HC}$ gauge interaction.
The SU(4)$_{ETC}$ gauge interaction and each of the descendents, SU(3)$_{ETC}$
and SU(2)$_{TC}$, as well as the SU(2)$_{HC}$, are asymptotically free.  A
choice of fermion content that can achieve the necessary two-stage breaking of
the SU(4)$_{ETC}$ symmetry includes the SM-nonsinglet fermions given in
Eqs. (\ref{quarks}) and (\ref{leptons}) with $N_{ETC}=4$, together with the
following SM-singlet fermion fields:
\beq
\psi_{i,R} \ : \quad (\bar 4,1,1,1)_{0,R}
\label{ngen2_psi}
\eeq
\beq
\chi^{i,\alpha}_R \ : \quad (4,2,1,1)_{0,R}
\label{ngen2_chi}
\eeq
and
\beq
\zeta^{ij,\alpha}_R \ : \quad (6,2,1,1)_{0,R} \ , 
\label{ngen2_zeta}
\eeq
where the numbers refer to the representations of ${\rm SU}(4)_{ETC} \times
{\rm SU}(2)_{HC} \times {\rm SU}(3)_c \times {\rm SU}(2)_L$ and the subscripts
denote the weak hypercharge.  The 6-dimensional representation of SU(4) is the
antisymmetric rank-2 tensor representation, $[2]_4 = \asym$, which is 
self-conjugate (and hence has zero gauge anomaly).

The first two coefficients of the SU(4)$_{ETC}$ beta function are $b_1 = 22/3$
and $b_2 = -109/12$. Nominally, Eq. (\ref{alfir}) would imply that the
perturbative two-loop beta function for the SU(4)$_{ETC}$ theory has an
infrared zero at $\alpha_{_{ETC,IR}} = 352\pi/109 \simeq 10$.  However, this is
so large that the perturbative beta function may not be reliable.  Fortunately,
the only result that we need concerning this beta function is reliable, namely
that the ETC coupling grows as the scale $\mu$ decreases.  As will be discussed
further below, the HC interaction will play the dominant role in condensate
formation. The SU(2)$_{HC}$ sector has 10 chiral fermions, or equivalently, 5
Dirac fermions, transforming according to the fundamental representation.
Using Eqs. (\ref{b1hc}) and (\ref{b2hc}) with $N_{f,1/2}=2N_{f,D,1/2}=10$, we
have, for the first two coefficients of the SU(2)$_{HC}$ beta function, $b_1 =
4$ and $b_2 = 9/2$. This perturbative HC beta function does not have a zero
away from the origin. Since the number $N_{f,D,1/2}=5$ is substantially less
than the estimate \cite{wtc2} for the critical value $N_{f,cr} \simeq 8$, we
can be confident that the HC interaction does, indeed, confine and break chiral
symmetry, as required for the ETC symmetry breaking.

\subsection{Condensation at $\Lambda_1$ Breaking SU(4)$_{ETC}$ to
  SU(3)$_{ETC}$}

To satisfy the requirement that the most attractive channels, in which the
fermion condensation occurs at the high scales above the EWSB scale, are not
those involving SM-nonsinglet, ETC-nonsinglet fermions, we again arrange the
initial conditions specifying the strengths of the gauge couplings in the
ultraviolet so that the HC gauge interaction is sufficiently stronger than the
ETC interaction that the most attractive channels are those involving
HC-nonsinglet (SM-singlet) fermions.  Given that one arranges the model in this
way, the most attractive channel for condensation is
\beq
(4,2,1,1)_0 \times (6,2,1,1)_0 \to (\bar 4,1,1,1)_0 \ , 
\label{ngen2_mac1}
\eeq
with $\Delta C_2 = C_2([2]_4) = 5/2$ for SU(4)$_{ETC}$ and $\Delta C_2 = 3/2$
for SU(2)$_{HC}$.  The associated condensate is 
\beq
\langle \epsilon_{ijk\ell}\epsilon_{\alpha\beta} \, 
\chi^{j,\alpha \ T}_R C \zeta^{k\ell,\beta}_R \rangle \ , 
\label{ngen2_mac1condensate}
\eeq
where $\epsilon_{ijk\ell}$ is the totally antisymmetric tensor density for
SU(4)$_{ETC}$.  This breaks ${\rm SU}(4)_{ETC}$ to ${\rm SU}(3)_{ETC}$ and is
invariant under SU(2)$_{HC}$. With no loss of generality, we may define the
uncontracted SU(4)$_{ETC}$ index in Eq. (\ref{ngen2_mac1condensate}) to be
$i=1$, so that this condensate is proportional to
\beqs
& & 
\langle \epsilon_{\alpha\beta}(
 \chi^{2,\alpha \ T}_R C \zeta^{34,\beta}_R 
-\chi^{3,\alpha \ T}_R C \zeta^{24,\beta}_R 
+\chi^{4,\alpha \ T}_R C \zeta^{23,\beta}_R) \rangle \ . \cr\cr 
& & 
\label{ngen2_mac1condensateform}
\eeqs
The six $\chi^{j,\alpha}_R$ with $j=2,3,4$, $\alpha=1,2$ and the six
$\zeta^{k\ell,\beta}_R$ with $k\ell = 34, \ 24, \ 23$ and $\beta=1,2$ involved
in this condensate gain dynamical masses of order $\Lambda_1$, and the seven
ETC gauge bosons in the coset ${\rm SU}(4)_{ETC}/{\rm SU}(3)_{ETC}$ gain
dynamical masses of order $g_{ETC}\Lambda_1$.  Note that the measure of
attractiveness for the channel in Eq. (\ref{ngen2_mac1}) with respect to the
SU(4)$_{ETC}$ interaction, $\Delta C_2= 5/2$, is less than the $\Delta C_2 =
15/4$ for the undesired condensation $4 \times \bar 4$ involving the left- and
right-handed chiral components of the HC-singlet ETC multiplets containing the
SM quarks and leptons (together with the respective techiquarks and
technileptons).  The model is constructed so that the HC gauge coupling at the
scale $\Lambda_1$ is sufficiently large that it overwhelms this difference in
$\Delta C_2$ values and makes Eq. (\ref{ngen2_mac1}) the most attractive
channel.

\subsection{Theory for $\Lambda_2 \le E < \Lambda_1$ and Condensation at 
$\Lambda_2$ Breaking SU(3)$_{ETC}$ to SU(2)$_{TC}$}

In the low-energy effective field theory operative at energy scales $\mu$
directly below $\Lambda_1$, the light fermions that are nonsinglets under the
strongly coupled SU(3)$_{ETC}$ and/or SU(2)$_{HC}$ gauge groups include the SM
nonsinglets in Eqs. (\ref{quarks}) and (\ref{leptons}) and the SM singlets (i)
$\zeta^{1j,\alpha}_R$, (ii) $\chi^{1,\alpha}_R$, and (iii) and $\psi_{j,R}$
with $j=2,3,4$.  These transform, respectively as (i) (3,2), (ii) (1,2), and
(iii) $(\bar 3,1)$ representations of ${\rm SU}(3)_{ETC} \times {\rm
SU}(2)_{HC}$.  As the theory evolves to lower energy scales $\mu$, the ETC and
HC gauge couplings continue to grow, and as $\mu$ decreases through a scale
that we denote $\Lambda_2$, the dominant SU(2)$_{HC}$ interaction, in
conjunction with the additional strong SU(3)$_{ETC}$ interaction, produces a
condensate in the most attractive channel, which is
\beq
(3,2,1,1)_0 \times (3,2,1,1)_0 \to (\bar 3,1,1,1)_0 \ . 
\label{ngen2_mac2}
\eeq
This has $\Delta C_2 = 4/3$ for SU(3)$_{ETC}$ and $\Delta C_2 = 3/2$ for
SU(2)$_{HC}$. The condensation in this channel breaks SU(3)$_{ETC}$ to
SU(2)$_{TC}$ and is invariant under SU(2)$_{HC}$. The associated condensate is
\beq
\langle \epsilon_{ijk}\epsilon_{\alpha\beta} \, 
\zeta^{1j,\alpha \ T}_R C \zeta^{1k,\beta}_R \rangle \ , 
\label{ngen2_mac2zetazeta}
\eeq
where $i,j,k \in \{2,3,4\}$.  With no loss of generality, we may choose $i=2$
as the breaking direction in SU(3)$_{ETC}$, so that this condensate takes 
the form
\beq
2\langle \epsilon_{\alpha\beta} \, 
\zeta^{13,\alpha \ T}_R C \zeta^{14,\beta}_R \rangle \ . 
\label{ngen2_mac2zetazetaform}
\eeq
The four chiral fermions $\zeta^{13,\alpha}_R$ and $\zeta^{14,\alpha}_R$ with
$\alpha=1,2$ gain masses of order $\Lambda_2$, and the five ETC gauge bosons in
the coset ${\rm SU}(3)_{ETC}/{\rm SU}(2)_{TC}$ gain masses of order $g_{_{ETC}}
\Lambda_2$.

\subsection{Condensation at $\Lambda_2' < \Lambda_2$}

The low-energy effective field theory operative just below $\Lambda_2$ is
thus invariant under the direct product group 
\beq
{\rm SU}(2)_{TC} \times {\rm SU}(2)_{HC} \times G_{SM} \ . 
\label{gtchcsm}
\eeq
The massless SM-singlet fermions that are nonsinglets under SU(2)$_{TC}$ or
SU(2)$_{HC}$ are $\chi^{1,\alpha}_R$, $\zeta^{12,\alpha}_R$, and $\psi_{i,R}$
with $i=3,4$. The first two of these transform as (1,2) under ${\rm SU}(2)_{TC}
\times {\rm SU}(2)_{HC}$ and the last as $(\bar 2,1) \approx (2,1)$.  Given
that $\alpha_{_{HC}} > \alpha_{_{TC}}$, the most attractive channel is $(1,2)
\times (1,2) \to (1,1)$, with condensate
\beq
\langle \epsilon_{\alpha\beta} 
\chi^{1,\alpha \ T}_R C \zeta^{12,\beta}_R \rangle \ , 
\label{ngen2_mac3chizeta}
\eeq
with $\Delta C_2=3/2$. The condensate is invariant under the full group
(\ref{gtchcsm}).  It forms at a scale $\Lambda_2' < \Lambda_2$, since it has
the same measure of attractiveness, $\Delta C_2$, with respect to SU(2)$_{HC}$
as the channel (\ref{ngen2_mac2}), but is driven by hypercolor alone, while the
channel (\ref{ngen2_mac2}) involves attraction due to both SU(2)$_{HC}$
and SU(3)$_{ETC}$.  The four chiral fermions $\chi^{1,\alpha}_R$ and 
$\zeta^{12,\alpha}_R$ get dynamical masses of order $\Lambda_2'$.

\subsection{Theory at Energy Scales Below $\Lambda_2'$} 

In the low-energy theory below $\Lambda_2'$, all of the HC-nonsinglet
fermions have gained dynamical masses and have consequently been integrated
out. The fermions in Eqs. (\ref{quarks}) and (\ref{leptons}) together with
$\psi_{i,R}$ with $i=3,4$ comprise 16 chiral doublets, or equivalently, 8 Dirac
doublets, of SU(2)$_{TC}$. As noted before, this TC theory
plausibly exhibits walking behavior.  The Dirac mass terms for the
technineutrinos are of the form (\ref{mnu_dirac_tc}).  The $\psi_{i,R}$ with
$i=1,2$ are TC-singlets that play the role of electroweak-singlet neutrinos.

Apart from mixing effects, the SM fermion masses of generation $i$ are 
generically of the form 
\beq
m_{f^i} \simeq \kappa \, \eta_i \, \frac{\Lambda_{TC}^3}{\Lambda_i^2} \ , \
i=1,2, 
\label{ngen2_mf}
\eeq
where the renormalization factor $\eta_i$ is given by 
\beq
\eta_i = \exp \left [ \int_{\Lambda_{TC}}^{\Lambda_i} \, \frac{d\mu}{\mu} \, 
\gamma(\alpha(\mu)) \right ] \ . 
\label{etai}
\eeq
Here, as before, $\alpha$ refers to the TC coupling as inherited from the
enveloping ETC theory.  (The generational index $i$ is written as an upper,
rather than lower, index on $f^i$ because the quarks and charged leptons arise
from fundamental, rather than conjugate fundamental, representations of the ETC
group.) Assuming that the theory exhibits walking up to $\Lambda_2'$, this
would be roughly of order
\beq
\eta_i \sim \frac{\Lambda_2'}{\Lambda_{TC}} 
\label{etaivalue}
\eeq
for both $i=1$ and $i=2$. It follows that, again neglecting mixing, 
\beq
m_{f^1} \simeq  \kappa \, \frac{\Lambda_2' \, \Lambda_{TC}^2}{\Lambda_1^2}
\label{mf1value}
\eeq
and
\beq
m_{f^2} \simeq  \kappa \, \frac{\Lambda_2' \, \Lambda_{TC}^2}{\Lambda_2^2} \ , 
\label{mfvalue}
\eeq
so that
\beq
\frac{m_{f^1}}{m_{f^2}} = \left( \frac{\Lambda_2}{\Lambda_1} \right )^2 \ . 
\label{mf12ratio}
\eeq
Thus, this model succeeds in producing a generational hierarchy in the
standard-model fermion masses.  Furthermore, because of the expected walking
behavior of the technicolor sector and the resultant enhancement of fermion
masses via the factor (\ref{etaivalue}), the fermions of the higher generation,
$i=2$, could have masses that are not too much smaller than the electroweak
symmetry breaking scale, $\Lambda_{TC}$.

\begin{figure}[hbtp]
\begin{center}
\includegraphics[3in, 8in][4in, 9.5in]{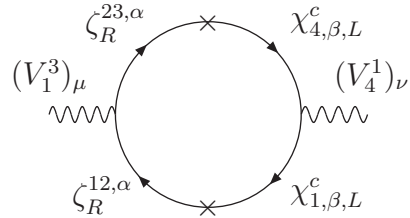}
\end{center}
\caption{\footnotesize{One-loop graph contributing to the ETC gauge boson
mixing $V_1^3 \leftrightarrow V_4^1$ (equivalently, 
       $V_1^4 \leftrightarrow V_3^1$) in the $N_{gen.}=2$ model.}}
\label{v1dn3uptov1up4dn}
\end{figure}

\begin{figure}[hbtp]
\begin{center}
\includegraphics[3in, 8in][4in, 9.5in]{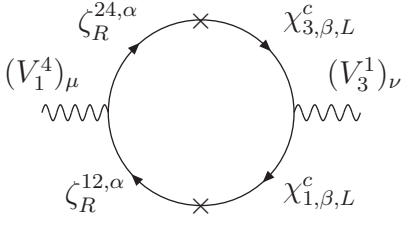}
\end{center}
\caption{\footnotesize{One-loop graph contributing to the ETC gauge boson
mixing $V_1^4 \leftrightarrow V_3^1$ (equivalently, 
       $V_1^3 \leftrightarrow V_4^1$) in the $N_{gen.}=2$ model.}}
\label{v1dn4uptov1up3dn}
\end{figure}

The various condensates of ETC-nonsinglet fermions give rise to corrections to
fermion propagators that are nondiagonal in ETC indices. In turn, via vacuum
polarization diagrams, these produce mixings of different ETC gauge bosons. In
Figs. \ref{v1dn3uptov1up4dn} and \ref{v1dn4uptov1up3dn} we show graphs
contributing to the ETC gauge boson mixing $V_1^3 \leftrightarrow V_4^1$, or
equivalently, $V_1^4 \leftrightarrow V_3^1$.  However, we find that the ETC
gauge boson mixing is not sufficient to give rise to the neutrino seesaw
mechanism of Ref. \cite{nt}, so one would have to add further ingredients to
the model in order to obtain appropriately small nonzero neutrino masses.

\section{A Model With $N_{gen.}=3$}
\label{model_ngen3}

\subsection{Field Content}

Here, for reference, we give a brief review of an $N_{gen.}=3$ model, which is
one of the models studied in Refs. \cite{nt,ckm,nag06}.  With $N_{gen.}=3$ and
$N_{TC}=2$, one uses SU(5)$_{ETC}$ for the ETC group.  The SM-nonsinglet
fermions and technifermions are given in Eqs. (\ref{quarks}) and
(\ref{leptons}).  The SM-singlet fermions are
\beq
\psi_{ij,R} \ : \quad (\overline{10},1,1,1)_{0,R}
\label{ngen3_psi}
\eeq
\beq
\zeta^{ij,\alpha}_R \ : \quad (10,2,1,1)_{0,R}
\label{ngen3_zeta}
\eeq
and
\beq
\omega_{\alpha,p,R} \ : \quad 2(1,2,1,1)_{0,R} \ , 
\label{ngen3_omega}
\eeq
where the first number in parentheses is the dimension of the representation of
SU(5)$_{ETC}$ and the others are the same as defined before.  In
Eq. (\ref{ngen3_omega}), $p=1,2$ is the copy number for the
$\omega_{\alpha,p,R}$ fields.  The $[2]_5$ is the rank-2 antisymmetric
representation of SU(5), with dimension 10. One includes an even number of
copies of the $\omega_{\alpha,p,R}$ field in order to avoid a global anomaly in
the SU(2)$_{HC}$ theory, and the choice of two copies is made to produce
desired mixings of ETC gauge bosons and resultant off-diagonal elements of
fermion mass matrices.

The SU(5)$_{ETC}$ beta function of this theory has leading coefficients $b_1 =
31/3$ and $b_2 = 224/15$.  Thus, the coupling $\alpha_{_{ETC}}$ increases to
large values as the energy scale decreases, triggering the formation of
condensates.  The SU(2)$_{HC}$ sector has 12 chiral fermions, or equivalently,
6 Dirac fermions, transforming according to the fundamental representation.
From Eqs. (\ref{b1hc}) and (\ref{b2hc}) with $N_{f,1/2}=2N_{f,D,1/2}=12$, it
follows that the first two coefficients of the SU(2)$_{HC}$ beta function are
$b_1 = 10/3$ and $b_2 = -11/3$. The number $N_{f,D,1/2}=6$ is less than the
estimated critical value of Dirac fermions, $N_{f,cr} \simeq 8$, leading to the
inference that the HC interaction confines and produces chiral condensates.

\subsection{Condensation at $\Lambda_1$ Breaking SU(5)$_{ETC}$ to
  SU(4)$_{ETC}$} 

In the context of variable $N_{gen.}$, one can comment on some features that
are present in this $N_{gen.}=3$ case that were not present for the lower
values of $N_{gen.}$.  Notably, for SU(5), fermions in the conjugate rank-2
antisymmetric $\overline{10}$ representation can play an important role in
self-breaking of the ETC symmetry.  In contrast, for SU(3) the rank-2
antisymmetric representation is equivalent to a conjugate fundamental
representation, while for SU(4) it is self-conjugate.  For SU(5), the
$\overline{10}$ can form a condensate with itself via the channel
\beq
(\overline{10},1,1,1)_0 \times (\overline{10},1,1,1)_0 \to (5,1,1,1)_0 \ . 
\label{10bar10barto5}
\eeq
The attractiveness of this channel, given by $\Delta C_2=24/5$, is the same as
for the undesired channel condensation (\ref{ffbar}). One can invoke vacuum
alignment arguments to infer that the initial condensation will occur in the
channel (\ref{10bar10barto5}) rather than (\ref{ffbar}). As was noted in
Ref. \cite{nt}, the channel (\ref{10bar10barto5}) is actually not the channel
with the largest value of $\Delta C_2$; the latter is $(\overline{10},1,1,1)_0
\times (10,2,1,1)_0 \to (1,2,1,1)_0$, with $\Delta C_2 = 36/5$, which would
leave SU(5)$_{ETC}$ invariant and would break SU(2)$_{HC}$.  One must thus
invoke a vacuum alignment and generalized most attractive channel argument to
infer that the latter condensation does not occur, since it would break the
strongly coupled HC interaction \cite{nt}. The formation of a condensate in the
channel (\ref{10bar10barto5}) breaks SU(5)$_{ETC}$ to SU(4)$_{ETC}$.  Hence, in
this $N_{gen.}=3$ case, in contrast to the situation for the $N_{gen.}=1,2$
models analyzed above in Sections \ref{model_ngen1} and \ref{model_ngen2}, one
can use the ETC interaction for the first stage of ETC gauge symmetry breaking.
Thus, here, the SU(5)$_{ETC}$ symmetry self-breaks, while in the $N_{gen.}=1,2$
models, the breaking of ETC is caused primarily by the HC interaction.  The
scale at which the condensate (\ref{10bar10barto5}) forms is denoted as
$\Lambda_1$. Choosing, as before, the direction of breaking to be $i=1$, one
obtains the condensate $\langle \epsilon^{1jk \ell n} \psi^T_{jk,R} C
\psi_{\ell n,R} \rangle$, or equivalently,
\beq
\langle \psi^T_{23,R} C \psi_{45,R} - \psi^T_{24,R} C \psi_{35,R} + 
\psi^T_{25,R} C \psi_{34,R} \rangle \ . 
\label{psipsicondensate}
\eeq
The six components $\psi_{jk,R}$ involved in this condensate gain dynamical
masses of order $\Lambda_1$.  The nine ETC gauge bosons in the coset 
${\rm SU}(5)_{ETC}/{\rm SU}(4)_{ETC}$ gain masses of order $g_{_{ETC}}\Lambda_1
\sim \Lambda_1$. The components of the multiplets in Eqs. (\ref{quarks}) and
(\ref{leptons}) with $i=1$ split off from the other components and become the
first generation of SM fermions.

\subsection{Theory for $\Lambda_2 \le E < \Lambda_1$ and Condensation at 
$\Lambda_2$ Breaking SU(4)$_{ETC}$ to SU(3)$_{ETC}$}

The low-energy effective field theory just below $\Lambda_1$ is invariant under
two strongly coupled gauge symmetries, SU(4)$_{ETC}$, acting on the ETC indices
$2 \le i \le 5$, and SU(2)$_{HC}$.  Decomposing the massless fermions inherited
from the SU(5)$_{ETC}$ theory in terms of representations of SU(4)$_{ETC}$ (and
the other exact symmetries at this level), one has the following content: (i)
$\psi_{1j,R}$, a $(\bar 4,1,1,1)_0$; (ii) $\zeta^{1j,\alpha}_R$, a
$(4,2,1,1)_0$; (iii) $\zeta^{jk,\alpha}_R$, a $(6,2,1,1)_0$; and (iii)
$\omega_{\alpha,p,R}$, forming two $(1,2,1,1)_0$ representations, where the
SU(4)$_{ETC}$ gauge indices are $2 \le i,j \le 5$, the SU(2)$_{HC}$ indices are
$\alpha=1,2$, and the copy index is $p=1,2$. The next two stages of ETC
symmetry breaking involve both the ETC and the HC interactions.  With the HC
interaction sufficiently strong, the next preferred step in gauge symmetry
breaking, occurring at the scale $\Lambda_2$, involves the formation of a
condensate in the most attractive channel
\beq
(4,2,1,1)_0 \times (6,2,1,1)_0 \to (\bar 4,1,1,1)_0 \ , 
\label{464barchannel}
\eeq
with $\Delta C_2 = 5/2$ for SU(4)$_{ETC}$ and $\Delta C_2=3/2$ for
SU(2)$_{HC}$.  This breaks SU(4)$_{ETC}$ to SU(3)$_{ETC}$ and preserves the
exact SU(2)$_{HC}$ symmetry.  Given that the SU(2)$_{HC}$ interaction is
strongly coupled at $\Lambda_2$, the HC glueballs are expected to have masses
of order $\kappa_g \Lambda_2$.

The symmetry breaking pattern in which this is the second stage was denoted
$G_a$ in Ref. \cite{nt} and sequence $S_1$ in Ref.  \cite{ckm}.  Note that,
with respect to the SU(4)$_{ETC}$ interaction, the value of $\Delta C_2$ for
this channel is less than the value $\Delta C_2=15/4$ for the undesired $4
\times \bar 4 \to 1$ channel (\ref{ffbar}) involving SM-nonsinglet
fermions. Thus, one again specifies a sufficiently large initial value for the
HC coupling $\alpha_{_{HC}}$ at a high scale so that the combination of the HC
and ETC interactions renders the channel (\ref{464barchannel}) more attractive
than the channel $4 \times \bar 4 \to 1$ channel. With no loss of generality,
one defines the index in which the SU(4)$_{ETC}$ breaks as $i=2$, so that the
condensate is
\begin{widetext}
\beq
\langle \epsilon_{\alpha\beta}\epsilon_{2jk \ell}\zeta^{1j,\alpha \ T}_R C
\zeta^{k \ell,\beta}_R \rangle  = 2 \langle \epsilon_{\alpha\beta}(
\zeta^{13,\alpha \ T}_R C \zeta^{45,\beta}_R -
\zeta^{14,\alpha \ T}_R C \zeta^{35,\beta}_R +
        \zeta^{15,\alpha \ T}_R C \zeta^{34,\beta}_R ) \rangle \ . 
\eeq
\end{widetext}
Here, $\epsilon_{ijk\ell}$ is the totally antisymmetric tensor density of the
SU(4)$_{ETC}$ theory resulting from the breaking of SU(5)$_{ETC}$ and hence is
identical to $\epsilon_{1ijk\ell}$ of the SU(5)$_{ETC}$ theory.  The twelve
$\zeta^{ij,\alpha}_R$ fields in this condensate gain masses of order
$\Lambda_2$, and the seven ETC gauge bosons in the coset ${\rm
SU}(4)_{ETC}/{\rm SU}(3)_{ETC}$ gain masses of order $g_{_{ETC}}\Lambda_2
\simeq \Lambda_2$.  At this scale $\Lambda_2$, the second generation SM
fermions, with $i=2$, split off from the other components of the multiplets in
Eqs. (\ref{quarks}) and (\ref{leptons}).

\subsection{Theory for $\Lambda_3 \le E < \Lambda_2$ and Condensation at 
$\Lambda_3$ Breaking SU(3)$_{ETC}$ to SU(2)$_{TC}$}

 Because the effective field theory below $\Lambda_2$ has SU(3)$_{ETC}$
symmetry (acting on the ETC indices $i=3,4,5$), to analyze this, we decompose
the SU(4)$_{ETC}$ representations in terms of SU(3)$_{ETC}$.  The massless
fermions that are nonsinglets under the ETC and/or HC groups operative here are
(i) $\psi_{1j,R}$, a $(\bar 3,1,1,1)_0$; (ii) $\zeta^{2j,\alpha}_R$, a
$(3,2,1,1)_0$ where $j=3,4,5$; (iii) $\zeta^{12,\alpha}_R$, a $(1,2,1,1)_0$;
and (iv) $\omega_{\alpha,p,R}$, comprising two $(1,2,1,1)_0$ representations.
Since the the SU(3)$_{ETC}$ and SU(2)$_{HC}$ interactions are asymptotically
free, $\alpha_{_{ETC}}$ and $\alpha_{_{HC}}$ continue to grow.  A third and
final stage of ETC symmetry breaking occurs at a scale denoted $\Lambda_3$.
Given the specification of the strengths of the ETC and HC interaction, the
most attractive channel is
\beq
(3,2,1,1)_{0,R} \times (3,2,1,1)_{0,R} \to (\bar 3,1,1,1)_0 \ , 
\eeq
with $\Delta C_2 = 4/3$ for SU(3)$_{ETC}$ and $\Delta C_2 = 3/2$ for
SU(2)$_{HC}$.  This breaks SU(3)$_{ETC}$ to SU(2)$_{TC}$ and preserves
SU(2)$_{HC}$.  With the breaking direction taken as $i=3$, the associated 
condensate is 
\beqs
& & \langle \epsilon_{3jk}\epsilon_{\alpha\beta}
\zeta^{2j,\alpha \ T}_R C \zeta^{2k,\beta}_R \rangle \ = \cr\cr
& & 
  \ 2 \langle \zeta^{24,1 \ T}_R C \zeta^{25,2}_R -
     \zeta^{24,2 \ T}_R C \zeta^{25,1}_R \rangle \ . 
\eeqs
Here, $\epsilon_{ijk}$ is the totally antisymmetric tensor density of
the SU(3)$_{ETC}$ theory resulting from the breaking of SU(4)$_{ETC}$ and 
hence is identical to $\epsilon_{2ijk}$ of the SU(4)$_{ETC}$ theory.
The six $\zeta^{2j,\alpha}_R$ fields involved in this condensate gain dynamical
masses of order $\Lambda_3$, and the five ETC gauge bosons in the coset ${\rm
SU}(3)_{ETC}/{\rm SU}(2)_{TC}$ gain masses of order $g_{_{ETC}}\Lambda_3 \simeq
\Lambda_3$.  At this scale $\Lambda_3$, the third generation of SM fermions
splits off, leaving the residual technifermions in each of the respective
multiplets.

\subsection{Condensation at $\Lambda_3' < \Lambda_3$}

Since the HC interaction is strong, it can also produce condensates involving
residual massless fermions that are singlets under SU(2)$_{TC}$ but 
nonsinglets under SU(2)$_{HC}$, in the channel 
\beq
(1,2,1,1)_0 \times (1,2,1,1) \to (1,1,1,1)_0 \ , 
\label{hc221}
\eeq
where the first number is the dimension of the representation with respect to
SU(3)$_{ETC}$ and the others are as before. The condensates that form in 
this channel include 
\beq
\langle \epsilon_{\alpha\beta} \zeta^{12,\alpha \ T}_R C
\zeta^{23,\beta}_R \rangle \ , 
\label{zeta12zeta23}
\eeq
\beq
\langle \epsilon_{\alpha\beta} \zeta^{12,\alpha \ T}_R C
\omega^\beta_{p,R} \rangle \ , 
\label{zeta12omega}
\eeq
\beq
\langle \epsilon_{\alpha\beta} \zeta^{23,\alpha \ T}_R C
\omega^\beta_{p,R} \rangle \ , 
\label{zeta34omega}
\eeq
and
\beq
\langle \epsilon_{\alpha\beta} \omega^{\alpha \ T}_{1,R} C
\omega^\beta_{2,R} \rangle \
\eeq
with $p=1,2$.  Since these condensates are formed only via the hypercolor
attraction, without any additional SU(3)$_{ETC}$ interaction, they form at a
scale $\Lambda_3' \lsim \Lambda_3$, where $\alpha_{_{HC}}$ has grown somewhat
larger than at $\Lambda_3$.  The fermions involved in these condensates get
dynamical masses of order $\Lambda_3'$.

\subsection{Theory at Energy Scales Below $\Lambda_3'$}

In the effective theory below $\Lambda_3$, all of the fermions
$\zeta^{ij,\alpha}_R$ have gained masses and have been integrated out, as have
all of the $\psi_{ij,R}$ for $2 \le i,j \le 5$, and all of the
$\omega_{\alpha,p,R}$ fields.  The resultant theory has one strongly coupled
symmetry with massless nonsinglet fermions, namely the technicolor group
SU(2)$_{TC}$.  The technifermions include those with SM-nonsinglet quantum
numbers, given in Eq. (\ref{quarks}) and (\ref{leptons}), and the $\psi_{1j,R}$
for $j=4,5$.  These constitute $4(N_c+1)=16$ chiral fermion doublets, or
equivalently, 8 Dirac fermion doublets, of SU(2)$_{TC}$.  Thus, $b_1=2$ and
$b_2=-20$ for the beta function of this theory, which has an approximate
infrared zero at $\alpha_{_{TC}} = 2\pi/5 \simeq 1.3$. To within the
theoretical uncertainties, this is equal to the critical value $\alpha_{cr}=
4\pi/9 \simeq 1.4$ for condensation of the technifermions to form the
condensates (\ref{technicondensate}).  Furthermore, as noted, since the number
of technifermions in this theory is close to $N_{f,cr}$, it plausibly exhibits
the desired property of walking associated with the approximate infrared fixed
point at $\alpha_{_{TC}}$.

The various condensates give rise to a variety of ETC gauge boson mixings.  In
turn, these lead to mass matrices for the quarks and charged leptons with both
hierarchical diagonal and off-diagonal entries.  The diagonal elements have the
generic form of Eq. (\ref{ngen2_mf}) with $i=1,2,3$.  Since the walking
behavior extends over the energy interval where there are eight massless Dirac
technifermions, namely from $\Lambda_{TC}$ to $\Lambda_3$, it follows that the
renormalization factor is roughly $\eta_i \sim \Lambda_3/\Lambda_{TC}$.  This
makes it possible for the mass of the top quark to be comparable to the
electroweak symmetry-breaking scale.  However, as noted, the model with the
content of quarks, leptons, techniquarks, and technileptons as given in
Eqs. (\ref{quarks}) and (\ref{leptons}) has difficulty explaining the large
splitting between $m_t$ and $m_b$. In general, the running masses of SM
fermions of the $i$'th generation, $m_{f^i}(p)$, are constants (apart from
logs) up to the highest scale, $\Lambda_i$, where they arise, and decay
asymptotically like $m_{f^i}(p) \propto \Lambda_i^2/p^2$ (up to logs) for $p >>
\Lambda_i$, where $p$ is a Euclidean momentum \cite{sml}. This model is also
able to produce appropriate Dirac and Majorana masses for neutrinos, in a
manner such as to yield a seesaw that generates acceptably light observed
neutrino mass eigenstates \cite{nt,ckm}.  Further details are given in
Refs. \cite{nt,ckm,kt,nag06}.

\section{A Model With $N_{gen.}=4$}
\label{model_ngen4}

\subsection{Field Content} 

One can also investigate a situation with $N_{gen.}$ larger than the inferred
physical value of 3 \cite{ngen3}.  In considering models with larger values of
$N_{gen.}$, we will require that these models retain the basic properties of
QCD, namely that (i) it is asymptotically free, which implies that the number
of quarks, $N_q$, is bounded above by $N_q < 33/2$; and (ii) as the scale
decreases below a GeV, QCD should confine and spontaneously break chiral
symmetry rather than evolving into the infrared in a chirally symmetric manner
such as would be associated with a non-Abelian Coulomb phase.  Analyses of the
Dyson-Schwinger equation for the quark propagator \cite{wtc2} yield $N_{f,cr}
\simeq 12$, and recent lattice simulations are broadly consistent with this
estimate, to within their theoretical uncertainties \cite{lgt}.  Keeping the
number of quarks below 12 means keeping the number of generations below 6,
which allows one to consider the values $N_{gen.}=4$ and $N_{gen.}=5$, given
the above constraints. Here we will study the case $N_{gen.}=4$.  As noted
before, although we will consider this case from the abstract field-theoretic
point of view of its effect in constructing a TC/ETC model, we note that there
are continuing studies of the possibility that there really are four
generations of SM fermions, which, however, must necessarily avoid having a
fourth light active neutrino \cite{fhs}-\cite{ngen4recent}.  Combining the
value $N_{gen.}=4$ with the value $N_{TC}=2$ in Eq. (\ref{netc}) yields
SU(6)$_{ETC}$ as the ETC gauge group.  Just as for $N_{gen.}=3$ one could use
purely ETC interactions for the first stage of ETC symmetry breaking (which is
thus self-breaking), so also for this $N_{gen.}=4$ case we find that one can
use the ETC interaction by itself for the first two stages of ETC
self-breaking, from SU(6)$_{ETC}$ to SU(5)$_{ETC}$ and then to SU(4)$_{ETC}$.
We rely on the hypercolor interaction to produce the final two stages of ETC
breaking down to SU(2)$_{TC}$. However, we specify the initial value of the HC
coupling slightly above the first condensation to be such that as the HC
coupling $\alpha_{_{HC}}$ grows, it becomes significantly large at the third
level of symmetry breaking, $\Lambda_3$.

We take the SM-singlet chiral fermions of the model to consist of 
\beq
\chi^i_R \ : \ ([1]_6,1,1,1)_0
\label{ngen4_chi}
\eeq
\beq
\psi^{ij}_R \ : \ ([2]_6,1,1,1)_0
\label{ngen4_psi}
\eeq
\beq
\eta^{ijk}_R \ : \ ([3]_6,1,1,1)_0
\label{ngen4_eta}
\eeq
\beq
\zeta_{i,\alpha,R} \ : ([\bar 1]_6,2,1,1)_0
\label{ngen4_zeta}
\eeq
and
\beq
\omega_{\alpha,p,R} \ : 2(1,2,1,1)_0 \ , 
\label{ngen4_omega}
\eeq
where the SU(6)$_{ETC}$ gauge indices run from 1 to 6, and the copy index on
$\omega_{\alpha,p,R}$ takes the values $p=1,2$. Note that since the
$\zeta_{i,\alpha,R}$ fields comprise an even number (six) of SU(2)$_{HC}$
doublets, it is necessary to use an even number of the ETC-singlet, HC-doublet
$\omega_{\alpha,p,R}$ fields to avoid a global SU(2) anomaly.  We use two
copies, $p=1,2$.  Since the dimensionality of the $[k]_N$ representation is ${N
\choose k}$, we have ${\rm dim}([1]_6)=6$, ${\rm dim}([2]_6)=15$, and ${\rm
dim}([3]_6)=20$.  By construction, this theory is free of anomalies in the
SU(6) gauged currents. The one- and two-loop coefficients of the SU(6)$_{ETC}$
beta function are
\beq
b_1 = \frac{38}{3} \ , \quad b_2 = \frac{76}{3}  \ . 
\label{ngen4_beta}
\eeq
Since these have the same sign, the SU(6)$_{ETC}$ beta function does not have a
perturbative zero away from the origin.  The SU(2)$_{HC}$ theory contains eight
chiral fermions, or equivalently four Dirac fermions, transforming as doublet
representations.  This is well below the estimated value $N_{f,cr} \simeq 8$
separating the chirally broken from chirally symmetric phases of an SU(2)
theory, so that we can be confident that the SU(2)$_{HC}$ interaction confines
and produces fermion condensates as required.

\subsection{Condensation at $\Lambda_1$ Breaking SU(6)$_{ETC}$ to 
SU(5)$_{ETC}$}

The most attractive channel and hence the one most likely for condensation to
form at the highest scale, is 
\beq
([2]_6,1,1,1)_0 \times ([3]_6,1,1,1)_0 \to ([\bar 1]_6,1,1,1)_0
\label{ngen4_mac1}
\eeq
with $\Delta C_2 = 7$, which breaks SU(6)$_{ETC}$ to SU(5)$_{ETC}$. As before,
we denote this first and highest ETC symmetry-breaking scale as $\Lambda_1$.
Note that this is more attractive than the undesired condensation channel $6
\times \bar 6 \to 1$ involving the left- and right-handed components of the
SM-nonsinglet fermions in Eqs. (\ref{quarks}) and (\ref{leptons}), which has
$\Delta C_2 = 35/6 = 5.83$. The condensate associated with the channel
(\ref{ngen4_mac1}) is
\beq
\langle \epsilon_{ijk\ell mn} \psi^{jk \ T}_R C \eta^{\ell m n}_R \rangle \ , 
\label{ngen4_mac1condensate}
\eeq
where $\epsilon_{ijk\ell mn}$ is the totally antisymmetric tensor density for
SU(6).  With no loss of generality, we can pick the uncontracted SU(6)$_{ETC}$
index to be $i=1$. The ${5 \choose 2}=10$ components $\psi^{jk}_R$ with
$2 \le j, \ k \le 6$ and $j \ne k$, and the ${5 \choose 3}=10$ components
$\eta^{\ell m n}_R$ with $2 \le \ell, \ m, \ n \le 6$ (with unequal values of
$\ell, m, n$) pick up dynamical masses of order $\Lambda_1$. The 11 ETC gauge
bosons in the coset ${\rm SU}(6)_{ETC}/{\rm SU}(5)_{ETC}$ also gain masses
$\sim g_{ETC} \Lambda_1 \sim \Lambda_1$.  At this stage, $\chi^1_R$ decouples
from the strong dynamics, since it is a singlet under the residual 
${\rm SU}(5)_{ETC} \times {\rm SU}(2)_{HC}$ interaction.

\subsection{Theory for $\Lambda_2 \le E < \Lambda_1$ and Condensation at 
$\Lambda_2$ Breaking SU(5)$_{ETC}$ to SU(4)$_{ETC}$}

The low-energy theory operative just below $\Lambda_1$ has two strongly
coupled gauge groups, SU(5)$_{ETC}$ and SU(2)$_{HC}$.  The content of massless
SM-singlet fermions that are nonsinglets under these groups, in addition to 
$\omega_{\alpha,R}$, is 
\beq
\chi^j_R \ : \ (5,1,1,1)_0
\label{ngen4_su5_chi}
\eeq
\beq
\psi^{1j}_R \ : \ (5,1,1,1)_0
\label{ngen4_su5_psi}
\eeq
and
\beq
\zeta_{j,\alpha,R} \ : \ (\bar 5,2,1,1)_0
\label{ngen4_su5_zeta}
\eeq
with $2 \le j \le 6$, and 
\beq
\eta^{1jk}_R \ : \ (10,1,1,1)_0 \ , 
\label{ngen4_su5_eta}
\eeq
where $2 \le j \ne k \le 6$.  A most attractive channel in this theory is
\beq
(10,1,1,1)_0 \times (10,1,1,1)_0 \to (\bar 5,1,1,1)_0
\label{1010to5bar}
\eeq
with $\Delta C_2=24/5$, breaking SU(5)$_{ETC}$ to SU(4)$_{ETC}$. With no loss
of generality, we may take the breaking direction in SU(5)$_{ETC}$ to be $i=2$,
so that associated condensate is
\beq
\langle \epsilon_{2jk\ell m} \eta^{1jk \ T}_R C \eta^{1\ell m}_R\rangle \ , 
\label{ngen4_su5_mac2_condensate}
\eeq
where $\epsilon_{ijk\ell m}$ is the totally antisymmetric tensor density of
SU(5)$_{ETC}$ and the sums over repeated indices are over the values 3, 4, 5,
and 6.  Since the measure of attractiveness for this condensation channel,
$\Delta C_2 = 24/5$, is less than the $\Delta C_2=7$ for the first
condensation, it follows that the scale, $\Lambda_2$, at which the second
condensation occurs is lower than the first condensation scale, $\Lambda_1$.
As in the $N_{gen.}=3$ case, one can use vacuum alignment arguments to infer
that the condensation (\ref{1010to5bar}) occurs instead of the unwanted
condensation (\ref{ffbar}) involving the SM-nonsinglet fermions (which has the
same value of $\Delta C_2=24/5$).  The six chiral fermions $\eta^{1jk}_R$ with
$3 \le j,k \le 6$ involved in the condensate (\ref{ngen4_su5_mac2_condensate})
gain dynamical masses of order $\Lambda_2$, and the nine ETC gauge bosons in
the coset ${\rm SU}(5)/{\rm SU}(4)_{ETC}$ gain masses of order
$g_{_{ETC}}\Lambda_2 \sim \Lambda_2$.  At this stage, $\chi^2_R$ and
$\psi^{12}_R$ decouple from the strong dynamics, since they are singlets under
the residual ${\rm SU}(4)_{ETC} \times {\rm SU}(2)_{HC}$ interaction.  Note
that, as far as SU(5)$_{ETC}$ is concerned, the channel $(10,2,1,1)_0 \times
(10,2,1,1)_0 \to (\bar 5,3,1,1)_0$ is as attractive as the channel
(\ref{1010to5bar}), with $\Delta C_2=24/5$.  However, since this involves a
symmetric combination of the two representations, it yields a triplet
representation under SU(2)$_{HC}$ and is thus repulsive with respect to the HC
interaction, with $\Delta C_2=-1/2$.  For this reason, one may safely conclude
that there is no condensation in this channel.

\subsection{Theory for $\Lambda_3 \le E < \Lambda_2$ and Condensation at 
$\Lambda_3$ Breaking SU(4)$_{ETC}$ to SU(3)$_{ETC}$}

The low-energy effective theory operative just below the scale $\Lambda_2$ is
invariant under two strongly coupled groups, ${\rm SU}(4)_{ETC} \times {\rm
SU}(2)_{HC}$, where the SU(4)$_{ETC}$ acts on the indices $3 \le j \le 6$.  The
content of massless fermions that are nonsinglets under this direct product
group includes 
\beq
\chi^j_R \ : \ (4,1,1,1)_0
\label{ngen4_su4_chi}
\eeq
\beq
\psi^{1j}_R \ : \ (4,1,1,1)_0
\label{ngen4_su4_psi}
\eeq
\beq
\zeta_{j,\alpha,R} \ : \ (\bar 4,2,1,1)_0
\label{ngen4_su4_zeta}
\eeq
and
\beq
\eta^{12j}_R \ : \ (4,1,1,1)_0 \ , 
\label{ngen4_su4_eta}
\eeq
with $3 \le j \le 6$, There are also massless fermions that are singlets under
SU(4)$_{ETC}$ and doublets under SU(2)$_{HC}$ namely $\zeta_{j,\alpha,R}$,
$j=1,2$ and $\omega_{\alpha,p,R}$.  Since the hypercolor interaction is
asymptotically free, its coupling, $\alpha_{_{HC}}$, increases as the reference
scale $\mu$ decreases from $\Lambda_1$ through $\Lambda_2$, and the initial
conditions can be chosen so that at the scale $\Lambda_3 < \Lambda_2$, the
SU(2)$_{HC}$ interaction is sufficiently strong to produce condensation of
HC-doublet fermions. The most attractive channel is 
\beq
(\bar 4,2,1,1)_0 \times (1,2,1,1)_0 \to (\bar 4,1,1,1)_0 \ , 
\label{ngen4_zetaomega}
\eeq
with $\Delta C_2=3/2$ for SU(3)$_{HC}$. This breaks SU(4)$_{ETC}$ to
SU(3)$_{ETC}$ and preserves SU(2)$_{HC}$. The associated condensate is
\beq
\langle \epsilon^{\alpha\beta}\zeta_{j,\alpha,R}^T C \omega_{p,\beta,R}\rangle 
\ . 
\label{ngen4_zeta3omega}
\eeq
With no loss of generality, we choose the ETC gauge index $j=3$ so that the
residual SU(3)$_{ETC}$ gauge symmetry acts on the indices 4, 5, and 6. We may
also choose the copy index to be $p=1$ for the $\omega_{\alpha,p,R}$ field.
The explicit condensate is then
\beq
\langle \epsilon^{\alpha\beta}\zeta_{3,\alpha,R}^T C \omega_{1,\beta,R}\rangle
\ . 
\label{ngen4_zeta3omega_explicit}
\eeq
The $\zeta_{3,\alpha,R}$ and $\omega_{1,\beta,R}$ fields pick up dynamical
masses of order $\Lambda_3$ and the ETC five gauge bosons in the coset ${\rm
  SU}(3)_{ETC}/{\rm SU}(2)_{TC}$ gain masses of order $g_{_{ETC}}\Lambda_3
\sim \Lambda_3$.  

With the same degree of attractiveness, and hence at the same scale,
$\Lambda_3$, there is an HC-driven condensation of two 
SU(4)$_{ETC}$-singlet fields in the channel $(1,2,1,1)_0 \times
(1,2,1,1)_0 \to (1,1,1,1)_0$. The associated condensate is
\beq
\langle 
\epsilon^{\alpha\beta}\zeta_{1,\alpha,R}^T C \zeta_{2,\beta,R}\rangle \ . 
\label{ngen4_zeta1zeta2}
\eeq
This is invariant under the same strongly coupled ${\rm SU}(3)_{ETC} \times 
{\rm SU}(2)_{HC}$ symmetry group as the condensate
(\ref{ngen4_zeta3omega}).  As a consequence of these condensations, the
fermions $\zeta_{j,\alpha,R}$ with $j=1,2,3$ and $\omega_{\alpha,R}$ gain
dynamical masses of order $\Lambda_3$ and the seven ETC gauge bosons in the
coset ${\rm SU}(4)_{ETC}/{\rm SU}(3)_{ETC}$ gain masses of order
$g_{_{ETC}}\Lambda_3 \sim \Lambda_3$.  At this stage, $\chi^3_R$,
$\psi^{13}_R$, and $\eta^{123}_R$ decouple from the strong dynamics since they
are singlets under the residual ${\rm SU}(3)_{ETC} \times {\rm SU}(2)_{ETC}$
interaction.

\subsection{Theory for $\Lambda_4 \le E < \Lambda_3$ and Condensation at 
$\Lambda_4$ Breaking SU(3)$_{ETC}$ to SU(2)$_{TC}$}

The effective field theory operative just below $\Lambda_3$ is invariant under
the strongly coupled group ${\rm SU}(3)_{ETC} \times {\rm SU}(2)_{HC}$, with
the SU(3)$_{ETC}$ acting on the indices $j=4, 5$, and 6.  The 
massless fermions that are nonsinglets under this group are (i)
$\chi^j_R$, $\psi^{1j}_R$, and $\eta^{12j}_R$, forming $(3,1,1,1)_0$
representations; (ii) $\zeta_{j,\alpha,R}$, forming $(\bar 3,2,1,1)_0$; and 
(iii) $\omega_{\alpha,2,R}$, forming $(1,2,1,1)_0$. The most attractive 
channel, which involves both SU(3)$_{ETC}$ and SU(2)$_{HC}$ interactions, is 
\beq
(\bar 3,2,1,1)_0 \times (\bar 3,2,1,1)_0 \to (3,1,1,1)_0 \ , 
\label{ngen4_3bar3barto3}
\eeq
with $\Delta C_2 = 4/3$ for SU(3)$_{ETC}$ and the usual $\Delta C_2=3/2$ for
SU(2)$_{HC}$.  This condensation breaks SU(3)$_{ETC}$ to SU(2)$_{TC}$ and
preserves the SU(2)$_{HC}$ symmetry. The associated condensate is 
$\langle \epsilon^{\alpha\beta}\epsilon^{ijk}\zeta_{j,\alpha,R}^T C 
\zeta_{k,\beta,R} \rangle$.  With no loss of generality, we may choose $i=3$ as
breaking direction in SU(3)$_{ETC}$, so that the actual condensate is
proportional to 
\beq
\langle \epsilon^{\alpha\beta}\zeta_{5,\alpha,R}^T C \zeta_{6,\beta,R}\rangle \
. 
\label{ngen4_zetazetalowest}
\eeq
We denote the energy scale at which this condensation occurs as $\Lambda_4$.
The $\zeta_{j,\alpha,R}$ with $j=5,6$ involved in this condensate gain
dynamical masses of order $\Lambda_4$, and the five ETC gauge bosons in the
coset ${\rm SU}(3)_{ETC}/{\rm SU}(2)_{TC}$ gain masses of order
$g_{_{ETC}}\Lambda_4 \simeq \Lambda_4$.  At this final stage of ETC symmetry
breaking, $\chi^4_R$, $\psi^{14}_R$, and $\eta^{124}_R$ decouple from the
strong dynamics, since they are singlets under the residual ${\rm SU}(2)_{TC}
\times {\rm SU}(2)_{HC}$ interaction.

\subsection{Condensation at $\Lambda_4' < \Lambda_4$}

The HC interaction can also produce a condensate involving fermions that are
SU(2)$_{TC}$ singlets. Since the formation of this condensate is not
aided by the SU(3)$_{ETC}$ interaction, it takes place at a somewhat lower 
scale than $\Lambda_4$, where $\alpha_{_{HC}}$ has grown to a somewhat larger
value.  We denote this scale as $\Lambda_4'$. The associated condensate is 
\beq
\langle \epsilon^{\alpha\beta}\zeta_{4,\alpha,R}^T C 
\omega_{\beta,2,R}\rangle \ . 
\label{zeta4_omega2}
\eeq
This condensate is invariant under the same strongly coupled symmetry group,
${\rm SU}_{TC} \times {\rm SU}(2)_{HC}$, as the condensate
(\ref{ngen4_zetazetalowest}).  The $\zeta_{4,\alpha,R}$ and
$\omega_{\beta,2,R}$ fermions get dynamical masses of order $\Lambda_4'$ due to
this condensation.  Thus, as the theory evolves below $\Lambda_4'$, all of the
HC-nonsinglet fermions have gained masses and have accordingly been integrated
out.

\subsection{Theory Below the Energy Scale $\Lambda_4$}

The theory below $\Lambda_4$ is invariant under the strongly coupled groups
${\rm SU}(2)_{TC} \times {\rm SU}(2)_{HC}$ and under $G_{SM}$, which is still
weakly coupled at this scale.  The SU(2)$_{TC}$ acts on the two remaining
unbroken ETC indices $j=5,6$. This SU(2)$_{TC}$ sector includes the 15
SM-nonsinglet techniquarks and technileptons in Eqs. (\ref{quarks}) and
(\ref{leptons}), together with three SM-singlet technifermions, $\chi^j$,
$\psi^{1j}_R$, and $\eta^{12j}_R$, with $j=5,6$, to make a total of 18 chiral
doublets, or equivalently, nine Dirac doublets.  If the technicolor theory
confines and produces the requisite bilinear technifermion condensates,
breaking electroweak symmetry, then this may be an acceptable illustrative 
model of a theory with $N_{gen.}=4$ SM generations.  However, we note the same
concern that was we mentioned earlier, namely that an SU(2)$_{TC}$ theory with
nine Dirac technifermions doublets might evolve into the infrared without
producing technifermion condensates and breaking electroweak symmetry.  It
would be very desirable to use lattice simulations to elucidate the boundary of
the chirally symmetric phase of SU(2) as a function of the content of light
fermions and to check the Dyson-Schwinger prediction of $N_{f,cr} \simeq 8$ for
fermions in the doublet representation. These would constitute a natural
extension of the intensive recent lattice work that has been performed for
SU(3) \cite{lgt}.

The sequential ETC breakings as discussed above would produce, as desired, a
hierarchy of SM fermion masses, with the diagonal elements of the respective
mass matrices given by the generic formula (\ref{ngen2_mf}), with $i=1,2,3,4$,
i.e., for the four generations. Assuming that the SU(2)$_{TC}$ sector would
confine, it would exhibit strong walking behavior, since the value of $N_f$ is
so close to the boundary with the chirally symmetric phase.  Hence, the
renormalization factor for the fermion bilinears would be $\eta_i \sim
\Lambda_4/\Lambda_{TC}$.  One could also study the various ETC gauge boson
mixings and resultant off-diagonal elements of fermion mass matrices, as well
as neutrino masses, especially the requirement of avoiding a fourth light
neutrino.  However, our results in this section above already demonstrate that
one can construct a plausibly tenable model with dynamical EWSB and four SM
fermion generations with a corresponding hierarchy of masses.

\section{Discussion and Conclusions}
\label{conclusions}

The origin of electroweak symmetry breaking and of the standard-model fermion
generations is an outstanding question in particle physics, and it is not yet
understood why $N_{gen.}=3$, rather than some other number.  In contrast to the
standard model, supersymmetric extensions thereof, and grand unified theories,
where one just puts the number $N_{gen.}$ in by hand in a manner that is
independent of the gauge group, this number plays a central role in the
structure and properties of an extended technicolor model, since it determines
what the initial ETC gauge symmetry is, via Eq. (\ref{netc}), and how many
stages of breaking the ETC symmetry undergoes as it is reduced to the TC
subgroup symmetry.  In this paper we have taken $N_{gen.}$ as a variable, and
have explored the consequences of varying this number, in the context of models
with dynamical EWSB.  We have explicitly demonstrated that one can construct
TC/ETC models with $N_{gen.}=1$, 2, and 4, extending the extensive previous
work for the physical case of $N_{gen.}=3$. Our results show that the auxiliary
strongly coupled gauge symmetry (hypercolor) is quite useful for obtaining the
desired ETC symmetry breaking for these cases $N_{gen.}=1, 2$, and 4, just as
it was for $N_{gen.}=3$. We have also shown how, for values of $N_{gen.}$ other
than 3, one can construct TC/ETC models in which the technicolor theory that
results from the sequential ETC symmetry breaking produces the necessary
technifermion condensates and plausibly exhibits the desired property of a
large but slowly running gauge coupling associated with an approximate
infrared-stable fixed point.  We have demonstrated that one can build TC/ETC
models that can yield generational hierarchies for all of the values of
$N_{gen.}$ that we considered.  Furthermore, because in each case we were able
to obtain a residual technicolor sector that can exhibit walking behavior and
hence enhancement of SM fermion masses, the fermions of the highest generation
generically have masses that can be comparable in size to the electroweak
breaking scale.  Stated in other terms, the real-world fact that the top quark
has a mass of order the EWSB scale could be shared by fermions of the highest
generation in these TC/ETC models with values of $N_{gen.}$ different from
3. It is interesting to compare this result with the situation with the
conventional Yukawa mechanism for producing SM fermion masses, where the
triviality property of the Yukawa interaction places an upper limit on the
Yukawa coupling and hence on the resultant fermion mass.  This triviality upper
limit on the fermion mass produced by the Yukawa coupling is also comparable to
the electroweak symmetry breaking scale, as has been shown by fully
nonperturbative, dynamical-fermion lattice simulations \cite{y}.  In a theory
with strong walking behavior, the effects of SM gauge couplings, which are
relatively small perturbations at the TeV scale, could be magnified.  However,
it is questionable whether the models would produce large intragenerational
mass splittings, in particular, between the charge 2/3 and charge $-1/3$ quarks
of a given generation.  Our results for $N_{gen.}=4$ may be useful for those
studying the possibility of a real fourth generation.  With an illustrative
example specifically constructed for the purpose in Sect. \ref{failure}, we
have also illustrated a problem that one can encounter in model-building, in
which an excessive number of technifermions can lead to a chirally symmetric
evolution of the (asymptotically free) technicolor theory rather than the
requisite formation of technifermion condensates at the electroweak scale.

Clearly, TC/ETC theories are subject to a number of severe phenomenological
constraints, and one does not yet know if the origin of electroweak symmetry
breaking is dynamical, or is due to the vacuum expectation value of a
fundamental Higgs field, as hypothesized in the standard model and
supersymmetric extensions thereof.  However, we believe that the present study
of models with variable $N_{gen.}$ yields useful insights into the role of this
number in theories with dynamical electroweak symmetry breaking and can be of
value in the continuing quest to understand the origin of standard-model
fermion generations.  Moreover, since this work involves analyses of patterns
of dynamical symmetry breaking of strongly coupled gauge theories, it is also
of more abstract field-theoretic interest in its own right. One looks forward
eagerly to the elucidation of the physics that is responsible for electroweak
symmetry breaking, and an answer to the question of whether it involves
strongly or weakly coupled interactions at the TeV scale, that will be
forthcoming soon from the Large Hadron Collider.

\bigskip
\bigskip
\bigskip

Acknowledgments: This research was partially supported by the grant
NSF-PHY-06-53342.

\section{Appendix}
\label{appendix} 

In this appendix we list some formulas that are relevant to our study of the
evolution of the TC and HC gauge interactions as functions of energy scale. We
consider a (zero-temperature) vectorial SU($N$) gauge group with $N_f$ massless
Dirac fermions in the fundamental representation.  It is assumed that this
theory is asymptotically free, i.e., $b_1 > 0$ in Eq. (\ref{beta}).  Let us
define
\beq
N_{f,IR} = \frac{34N^3}{13N^2-3} \ , 
\label{nfir} 
\eeq
If $N_f < N_{f,I}$, then $b_2 > 0$, and the only zero of the perturbative
two-loop beta function is the zero at the origin (the ultraviolet fixed point
of the renormalization group).  As $N_f$ increases through the value
$N_{f,IR}$, $b_2$ reverses sign and becomes negative, so that 
the beta function has a zero away from the origin, at the value 
\beq
\alpha_{IR} = \frac{-4\pi(11N -2N_f)}{34N^2-13NN_f+3N^{-1}N_f} \ .
\label{alpha_irfp}
\eeq
As the energy scale $\mu$ decreases from large values, $\alpha$ increases
toward the value $\alpha_{IR}$, which is thus an infrared fixed point of the
renormalization group.  It is (i) an exact IR fixed point if there is no change
in the massless particle content as $\alpha$ increases toward $\alpha_{IR}$
from below, or alternatively (ii) an approximate IR fixed point if, as $\alpha$
increases toward $\alpha_{IR}$, it exceeds a critical value $\alpha_{cr}$ for
spontaneous chiral symmetry breaking via the formation of bilinear condensates
of the fermions at some scale $\mu = \Lambda_{cr}$.  In the latter case, these
fermions gain dynamical masses and are integrated out in the effective
low-energy theory that is applicable for $\mu < \Lambda_{cr}$; as a
consequence, the massless particle content of the theory changes and it evolves
further into the infrared in a manner governed by a different set of
coefficients in the beta function.

As $N_f$ increases above $N_{f,IR}$, the value of $\alpha_{IR}$ decreases, and
as $N_f$ increases through a critical value $N_{f,cr}$, $\alpha_{IR}$ decreases
below the minimum value, $\alpha_{cr}$, for condensate formation. For $N_{f,cr}
< N_f < (11/2)N$, the theory is therefore in a chirally symmetric phase.  In
accord with physical arguments connecting confinement and spontaneous chiral
symmetry breaking \cite{casher}, this is often inferred to be a conformal,
non-Abelian Coulomb phase.  This inference is clearly valid in the limit where
$N_f$ approaches $(11/2)N$ from below, so that $b_1$ and $\alpha_{IR}$ become
very small, and the gauge interaction become very weak.  The inference assumes
that in this phase without any fermion condensates, where the fermions do not
pick up any dynamical masses, there are also no glueballs; i.e., the glueballs
of the confined phase have become unbound, producing free massless gluons. The
value of $N_{f,cr}$ is determined by setting $\alpha_{IR} = \alpha_{cr}$,
yielding the result \cite{wtc2}
\beq
N_{f,cr} = \frac{2N(50N^2-33)}{5(5N^2-3)} \ .
\label{nfcr}
\eeq
For $N=2$, this gives $N_{f,cr} \simeq 8$ (and for $N=3$ it gives $N_{f,cr}
\simeq 12$).  This is the basis for the statement that a one-family technicolor
theory, which has $2(N_c+1)=8$ Dirac technifermions, plausibly exihibits
walking behavior.  Clearly, Eq. (\ref{alfcrit}) and the resultant
Eq. (\ref{nfcr}) are only rough estimates, in view of the strongly coupled
nature of the physics and the fact that this approach neglects nonperturbative
effects, such as instantons, which enhance chiral symmetry breaking \cite{as}.
Moreover, the Dyson-Schwinger equation does not incorporate confinement, and
the condition in Eq. (\ref{alfcrit}) is obtained by doing a loop integration
over all Euclidean loop momenta, but in fact the integration range is reduced,
since a particle confined within a size $r \sim 1/\Lambda$ has a maximum
wavelength and equivalently, a minimum boundstate momentum of order $\Lambda$,
the confinement energy scale \cite{lmax}.  Fortunately, these two omissions
(instantons and reduction of the integration range in the Dyson-Schwinger
integral) affect the prediction for $\alpha_{cr}$ in opposite ways, so that the
omission of both of them may not be too serious.  As noted in the text, a
continuum study of corrections to the one-gluon exchange approximation in
solving the Dyson-Schwinger equation found it to be reasonably accurate
\cite{alm}. More recently, in the case of SU(3), lattice studies yield results
that are broadly consistent with the predictions of the earlier Dyson-Schwinger
analysis \cite{lgt}.  In this context, one should note that the study of the
Dyson-Schwinger equation for the fermion propagator only gives information
about chiral symmetry breaking; this equation does not directly contain
information about confinement.  In principle, if appropriate conditions were
satisfied \cite{thooft} (which are necessary but not sufficient conditions),
one could have a confined phase without spontaneous chiral symmetry breaking.
However, this possibility is not relevant for our present analysis, since we
require that there be S$\chi$SB in the ETC and HC sectors to produce the
sequential ETC symmetry breaking, and in the TC sector to produce the
electroweak symmetry breaking. 

For our analyses of the successive stages of ETC symmetry breaking we will 
apply this sort of method for a chiral gauge theory with a general set of
fermion representations.  In this case, in terms of the chiral fermion
representations $R$, the first two coefficients of the beta function are 
\cite{b1} 
\beq
b_1 = \frac{1}{3}\bigg [ 11C_2(G)- 2 \sum_R T(R) N_{f,R} \bigg ]
\label{b1}
\eeq
and \cite{b2} 
\beq
b_2 = \frac{1}{3}\bigg [34C_2(G)^2 -
\sum_R [10 C_2(G) + 6C_2(R)] T(R)N_{f,R} \bigg ] \ . 
\label{b2}
\eeq
Higher coefficients are scheme-dependent, and these first two, which are
scheme-independent, will suffice for our purposes.  For a vectorial theory, the
left- and right-handed chiral fermions of representation $R$ are combined into
a single Dirac fermion in this representation. It is not necessary that the
enveloping ETC group or the intermediate subgroups above the level of
SU(2)$_{TC}$ exhibit walking behavior; our only constraints for these groups
are (i) that they be asymptotically free and (ii) that their fermion content be
such that as they evolve into the infrared, they spontaneously break chiral
symmetry via formation of fermion condensates instead of evolving into a
non-Abelian Coulomb phase. This does not require a perturbative infrared zero
of the beta function.

In particular, given that the SU(2)$_{HC}$ gauge interaction has an even
number, $N_{f,1/2}$, of chiral fermions transforming as doublet 
representations, as it must to avoid a global anomaly, it can always be
rewritten as a vectorial gauge theory with $N_{f,D,1/2}=N_{f,1/2}/2$ Dirac
doublets.  Hence for this HC theory we have 
\beq
b_1 = \frac{1}{3}(22-2N_{f,D,1/2})
\label{b1hc}
\eeq
and
\beq
b_2=\frac{1}{3}\left [ 136 - \frac{49 N_{f,D,1/2}}{2} \right ] \ .  
\label{b2hc}
\eeq


\begin{thebibliography}{99}
\bibitem{ngen3}
More precisely, for SM generations with associated neutrinos that are light
compared with $m_Z/2$, LEP has determined that $N_{\nu}=N_{gen.}=3$ from
measurements of the invisible width of the $Z$. This result is also consistent
with constraints on light neutrinos from primordial nucleosynthesis. 
Constraints on a possible fourth generation are noted in the text.
\bibitem{lambhc}
%
For our present purposes, it will be sufficient to restrict our consideration
to models in which the hypercolor symmetry is never broken.  We note, however,
that in one of the breaking sequences considered earlier, denoted $G_b$ in
Ref. \cite{nt} and sequence $S_2$ in Ref. \cite{ckm}, SU(2)$_{HC}$ is broken to
the (non-asymptotically free) U(1)$_{HC}$.
\bibitem{casimir}
%
The quadratic Casimir invariant for a group $G$ is given by
$\sum_{j=1}^{dim(R)} [D_R(T_a)]_{ij} [D_R(T_a)]_{jk} = C_2(R)\delta_{ik}$, 
where $a,b$ are group indices, $T_a$ are the generators of the associated Lie 
algebra, and $D_R(T_a)$ are the matrix representations of the $T_a$ in the
representation $R$.
\end{thebibliography}
\end{document}